\documentclass[11pt,letterpaper]{article}
\pdfoutput=1
\usepackage{jheppub}
\usepackage{color}
\usepackage{graphicx}
\usepackage{wrapfig}

\usepackage{verbatim}
\usepackage{amsmath}
\usepackage{amssymb}
\usepackage{subfig}
\usepackage{url}
\usepackage{bbold}
\usepackage{xspace}

\usepackage{multirow}
\usepackage{threeparttable}
\usepackage{paralist}

\usepackage{color}

\usepackage{floatrow}
\newfloatcommand{capbtabbox}{table}[][\FBwidth]

\DeclareRobustCommand{\Ref}[1]{Ref.~\cite{#1}}

\newcommand{\beq}{\begin{equation}}
\newcommand{\eeq}[1]{\label{#1}\end{equation}}
\def\beqa{\begin{eqnarray}}
\def\eeqa#1{\label{#1}\end{eqnarray}}
\newcommand{\eeqn}{\end{equation}}
\newcommand{\CR}{\notag \\}
\newcommand{\leqn}[1]{(\ref{#1})}
\def\nn{\nonumber}

\def\eg{\epsilon_\gamma}
\def\Ref{R_{\rm eff}}

\def\stacksymbols #1#2#3#4{\def\theguybelow{#2}
    \def\vp{\lower#3pt}
    \def\sp{\baselineskip0pt\lineskip#4pt}
    \mathrel{\mathpalette\intermediary#1}}

\def\intermediary#1#2{\vp\vbox{\sp
     \everycr={}\tabskip0pt
     \halign{$\mathsurround0pt#1\hfil##\hfil$\crcr#2\crcr
              \theguybelow\crcr}}}

\def\gsim{\stacksymbols{>}{\sim}{2.5}{.2}}
\def\lsim{\stacksymbols{<}{\sim}{2.5}{.2}}
\begin{document}

\title{Phenomenology of ELDER Dark Matter}

\author[a,b]{Eric Kuflik,}
\author[a]{Maxim Perelstein,}
\author[a,c]{Nicolas Rey-Le Lorier,}
\author[a,d]{and Yu-Dai Tsai}

\affiliation[a]{Laboratory for Elementary Particle Physics, Cornell University, Ithaca, NY 14853, USA}
\affiliation[b]{Racah Institute of Physics, Hebrew University of Jerusalem, Jerusalem 91904, Israel}
\affiliation[c]{Simpson College,
Indianola, IA 50125, USA}
\affiliation[d]{Perimeter Institute for Theoretical Physics, Waterloo, ON N2J 2W9, Canada}

\emailAdd{eak245@cornell.edu}
\emailAdd{mp325@cornell.edu}
\emailAdd{nicolas.reylelorier@simpson.edu}
\emailAdd{yt444@cornell.edu}

\date{\today}

\abstract{We explore the phenomenology of Elastically Decoupling Relic (ELDER) dark matter. ELDER is a thermal relic whose present density is determined primarily by the cross-section of its elastic scattering off Standard Model (SM) particles. Assuming that this scattering is mediated by a kinetically mixed dark photon, we argue that the ELDER scenario makes robust predictions for electron-recoil direct-detection experiments, as well as for dark photon searches. These predictions are independent of the details of interactions within the dark sector. Together with the closely related Strongly-Interacting Massive Particle (SIMP) scenario, the ELDER predictions provide a physically motivated, well-defined target region, which will be almost entirely accessible to the next generation of searches for sub-GeV dark matter and dark photons. We provide useful analytic approximations for various quantities of interest in the ELDER scenario, and discuss two simple renormalizable toy models which incorporate the required strong number-changing interactions among the ELDERs, as well as explicitly implement the coupling to electrons via the dark photon portal.}

\maketitle

\section{Introduction}

Cosmological observations at a variety of length scales, from individual galaxies to the Hubble scale, indicate that most of the matter in the universe is in the form of dark matter (DM). DM cannot consist of any of the known elementary particles, and its existence provides solid experimental evidence for physics beyond the Standard Model (SM). The microscopic nature of dark matter is one of the major mysteries in fundamental physics. For many years, both theoretical work and experimental searches for dark matter focused on a short list of possible candidates independently motivated by particle physics---primarily QCD axions and weakly-interacting massive particles (WIMPs) realized within supersymmetry or other extensions of the SM at the weak scale. Despite decades of experimental effort, no evidence for these candidates has been found. While neither WIMP nor axion dark matter is ruled out and the experimental searches are ongoing, there has been renewed interest in exploring alternative particle dark matter candidates. 

A promising new direction is to consider models in which dark matter particles have strong number-changing self-interactions~\cite{Carlson:1992fn,Hochberg:2014kqa,Bernal:2015bla,Lee:2015gsa,Bernal:2015ova,Bernal:2015xba,Kuflik:2015isi,Hochberg:2015vrg,Choi:2016hid,Pappadopulo:2016pkp,Farina:2016llk,Forestell:2016qhc,Soni:2016gzf,Choi:2016tkj,Cline:2017tka,Choi:2017mkk,Bernal:2017mqb,Dey:2016qgf}. If the DM is a thermal relic, its current density in such models can be determined either by the cross section of the number-changing self-interaction processes (``Strongly-Interacting Massive Particle", or SIMP, scenario~\cite{Hochberg:2014dra}) or by the cross section of elastic scattering between the DM and SM (``Elastically Decoupling Relic", or ELDER, scenario~\cite{Kuflik:2015isi}). In both cases, the observed DM density is naturally obtained if the mass of the DM particles is parametrically close to the QCD confinement scale, $m_{\rm DM} \sim 10-100$ MeV. This leads to an attractive particle physics framework: a ``dark sector" of fields not charged under the SM gauge groups, containing a non-Abelian ``dark QCD" gauge group that confines at a scale similar to $ \Lambda_{\rm QCD}$. The proximity of the SM and ``dark" confinement scales may be due to a discrete symmetry relating the dark QCD gauge coupling to the SM $g_3$ at a high energy scale~\cite{Foot:1991bp,Foot:1991py,Berezhiani:1995am,Chacko:2005pe,Burdman:2006tz}. The dark matter may then consist of mesons that emerge from dark QCD upon confinement~\cite{Hochberg:2014kqa}. If the dark sector also contains an Abelian gauge field, kinetic mixing between this field and the SM electromagnetic field naturally provides the requisite interaction between the dark matter particle and the SM, via the dark photon portal~\cite{Lee:2015gsa,Hochberg:2015vrg}. 

The goal of this paper is to study the above possibilities in more detail, in particular, the ELDER scenario proposed in Ref.~\cite{Kuflik:2015isi}. In Ref.~\cite{Kuflik:2015isi}, we demonstrated the viability of this scenario in a general framework, without reference to a specific model of either the dark sector or the portal connecting it to the SM. Instead, we used a simple parametrization of the DM number-changing self-scattering and DM-SM elastic scattering cross sections. Moreover, the analysis of Ref.~\cite{Kuflik:2015isi} was primarily based on numerical solution of Boltzmann equations. Here, we expand that analysis in several directions:

\begin{itemize}

\item We provide an approximate analytic solution to the Boltzmann equations that describe the evolution of the ELDER dark matter density during the epoch of its kinetic decoupling from the SM. This in turn leads to precise analytic estimates of ELDER relic density, and hence the model parameters required to obtain the observed dark matter abundance. We also combine these estimates with unitarity considerations to obtain a model-independent upper bound on the ELDER dark matter mass. This is the subject of Section~\ref{sec:ELDER}. 

\item We consider the phenomenology of the ELDER scenario with a dark photon portal mediating the interactions between the ELDERs and the SM. We find that the model makes a {\it remarkably robust} prediction for rates expected in direct-detection experiments. This prediction has no free parameters beyond the ELDER particle mass, and is completely independent of the details of the dark-sector self-interactions. The reason is that the ELDER relic density with this portal is determined by the cross section of elastic scattering of dark matter particles on electrons, which is precisely the same process used for direct detection in the MeV-GeV DM mass range. This feature is unique to ELDERs. Likewise, this scenario provides firm predictions for dark photon searches.  
Together with the well-known ``thermal targets", the ELDER and SIMP predictions define a well-defined target region in the parameter space for direct-detection and dark photon experiments, bounded from all sides. These findings are reported in Section~\ref{sec:pheno}. In particular, Figs.~\ref{fig:DD} and~\ref{fig:DP} encapsulate the main results of this paper.

\item We discuss two simple perturbative models for the dark sector, which realize the ELDER scenario with the dark photon portal; see Section~\ref{sec:scalars}. These can be thought of as toy models that describe interactions among low-lying mesons created by confining gauge dynamics in the gauge sector. 

\end{itemize}

Details of the Boltzmann equations, an approximate analytic solution for kinetic decoupling, and some useful formulas for thermally-averaged rates, are collected in the Appendices. 

\section{ELDER Dark Matter}
\label{sec:ELDER}

Consider a particle $\chi$ with mass $m_\chi$\footnote{This may be a single state, or a set of mass-degenerate states $\chi_i$. In the latter case, appropriate averaging over the particle ``flavor" is implicit in the discussion of this section, and the ``flavor indices" are suppressed for clarity.}. The $\chi$ particles can undergo the following processes:

\begin{enumerate}

\item Elastic scattering: $\chi +\rm{SM}\leftrightarrow \chi+\rm{SM}$, where ``SM" stands for any of the Standard Model particles. (In practice, the important SM states are those with mass below $m_\chi$; for ELDERs, this will typically include electrons, photons, and neutrinos.) 

\item Annihilations to SM: $\chi+\chi \leftrightarrow$ SM + SM. 

\item ``$3\to 2$" Self-Annihilations: $\chi\chi\chi\leftrightarrow \chi\chi$.  

\item ``$2\to 2$" Elastic Self-Scattering: $\chi\chi\leftrightarrow\chi\chi$. 

\end{enumerate}

We assume that in the early universe at temperatures above $m_\chi$, all four reactions are ``active", {\it i.e.} occur in the plasma at rates $\Gamma > H$. This means that the ELDERs have a thermal energy distribution (thanks to reaction 4), zero chemical potential (reaction 3), and temperature equal to that of the SM plasma (reactions 1 and 2), which we denote by $T$. The ELDER number density follows the equilibrium trajectory, $n^{\rm eq}(T)$. As the temperature drops below $m_\chi$, the ELDERs become non-relativistic, and the equilibrium density drops exponentially, $n^{\rm eq}(T)\propto e^{-m_\chi/T}$. The rates of the reactions 2, 3, and 4, drop off exponentially, while the reaction 1 slows more gradually. 

All the reactions eventually decouple, $\Gamma \lsim H$, but the order of decoupling is crucially important in determining the relic abundance. It is natural for $3\to 2$ self-annihilation to decouple before $2\to 2$ self-scattering: the interaction strengths entering the two rates are generically of the same order (both involve interactions internal to the dark sector), but $\Gamma_{3\to 2} \propto n_\chi^2$ while $\Gamma_{2\to 2}\propto n_\chi$. On the other hand, the rate of annihilations to SM, $\Gamma_{\rm an}$, is controlled by the coupling between the SM and the dark sector, which can naturally be small. (For example, in the dark photon portal model considered below, this will be controlled by kinetic mixing between the SM and dark-sector $U(1)$ gauge groups.) In this paper, we will consider the regime where annihilations to SM decouple {\it first}, while the $3\to 2$ process is still active. This is the case in both the SIMP and ELDER scenarios.

The rate of elastic scattering $\Gamma_{\rm el}$ is proportional to the SM density, which is not exponentially suppressed at $T<m_\chi$. However, the scattering cross section is suppressed by the small coupling between the SM and $\chi$. Generically, this cross section is of the same order as that of annihilations to SM, and therefore decoupling of elastic scattering occurs after annihilations to SM are decoupled.  Depending on the relative strength of the SM-$\chi$ coupling and $\chi$ self-couplings, the decoupling of the elastic scattering may occur either after or before the decoupling of the $3\to 2$ self-annihilation. The former case corresponds to the SIMP scenario~\cite{Hochberg:2014dra}, while the latter is the ELDER scenario~\cite{Kuflik:2015isi}. 

\subsection{The Thermal History of ELDERs}

After annihilations and elastic scattering with the SM decouple, but while the $3\to2$ and $2\to2$ self-interactions are still active, the ELDERs are still in thermal equilibrium at zero chemical potential, but their temperature $T^\prime$ no longer has to be the same as the SM plasma temperature $T$. As shown in Appendix B, the two temperatures are related by 
\beq
\frac{\partial T^\prime}{\partial T} = 3 \frac{T^{\prime\,2}}{m_\chi T} +a\left(\frac{T}{m_\chi}\right)^{1+n} \frac{T^{\prime\,2}}{m_\chi^2}\frac{(T^\prime -T)}{m_\chi},
\eeq{TprimeAndT}
where
\beq
a \equiv  \frac{c_n g_\psi^2 g_\chi N^\psi_{3+n}}{32 \pi^3} \,\frac{M_{\rm Pl}}{1.66 g_{*,d}^{1/2} m_\chi}.
\eeq{adef}
Here $\psi$ is the SM particle that couples to $\chi$, with corresponding number of degrees-of-freedom $g_\psi$ and $g_\chi$, respectively; $N_{3+n}^\psi$ is a numerical constant given in Eq.~\leqn{Ndef}. We assume that the effective number of relativistic degrees of freedom $g_{*,d}$ remains constant throughout the decoupling process. (The case of varying $g_*$ can be handled numerically.) The {\it ``elastic scattering strength"} $c_n$ is defined as the dimensionless coefficient of the leading term in the low-energy expansion of the matrix element-squared of the elastic scattering process $\chi\psi\leftrightarrow\chi\psi$:
\beq
  \mathop{\hspace{-6.5ex}\left|\mathcal{M}\right|^2_{t=0}}_{\hspace{6.5ex}s=m_\chi^2+2m_\chi E_\psi}
 \equiv  c_n \left(\frac{E_{\psi}}{m_\chi}\right)^{n} + \ldots, 
\eeq{cndef}
where $|\mathcal{M}|^2$ is averaged over initial and final-state degrees of freedom, including spin, color, and electric charge. (See Appendix A for details.) If $\chi$ couples to more than one SM particle, a summation over the relevant SM species is implied in the definition of $a$. The formalism presented here is applicable to SM particles that are relativistic at the time of $\chi$ decoupling, $m_{\rm SM} \ll T_d\sim m_\chi/10$. SM particles with $m_{\rm SM} \gg T_d$ are irrelevant to the decoupling process, while the case $m_{\rm SM} \sim T_d$ can be studied numerically.

An approximate analytic solution to the temperature evolution equation can be found (see Appendix B):
\beq
x^\prime = 
e^t \left(\left(\frac{a}{n+4}\right)^{\frac{1}{n+4}} \Gamma \left(\frac{n+3}{n+4},t\right)-\frac{3 \text{Ei}(-t)}{n+4}\right)\,,
\eeq{ANAL}
where $x=m_\chi/T$, $x^\prime=m_\chi/T^\prime$, and $t= \frac{a x^{-n-4}}{n+4}$. At small $x$, $x^\prime\approx x$, corresponding to SM and ELDER sectors in thermal equilibrium. At large $x$, the asymptotic form of the solution is
\beq
x^\prime \approx   3 \log (x)+\left(\frac{a}{n+4}\right)^{\frac{1}{n+4}} \Gamma \left(\frac{n+3}{n+4}\right)-3 \log \left[e^{\frac{\gamma_E }{n+4}} \left(\frac{a}{n+4}\right)^{\frac{1}{n+4}}  \right].
\eeq{asymp}
Identifying the ``decoupling temperature" at which the ELDER and the SM thermally decouple, 
\beq
T_d = m_\chi \frac{\left(\frac{n+4}{a}\right)^{\frac{1}{n+4}}}{\Gamma \left(\frac{n+3}{n+4}\right)},
\eeq{Td}
Eq.~\leqn{asymp} can be rewritten as $x^\prime \simeq x_d + 3\log (x/x_d)$, or  
\beq
T^\prime \simeq \frac{T_d}{1 + 3 \frac{T_d}{m_\chi} \log \frac{T_d}{T} }.
\eeq{cannib}
This is precisely the behavior expected in the ``cannibalization" regime~\cite{Carlson:1992fn}, where ELDER temperature decreases only slowly (logarithmically with the scale factor) as the universe expands. The physical reason is that the kinetic energy released by $3\to 2$ self-annihilations partially compensates for the energy lost when particle momenta are redshifted due to the expansion. This regime persists until the $3\to 2$ process decouples, after which the ELDER density is frozen out. Note that the dark matter particles remain non-relativistic throughout the cannibalization period, so that from the point of view of Cosmic Microwave Background (CMB) and structure formation, ELDER is a Cold Dark Matter (CDM) candidate, consistent with observations.

\begin{figure}[t!]
\begin{center}
\includegraphics[width=8.7cm]{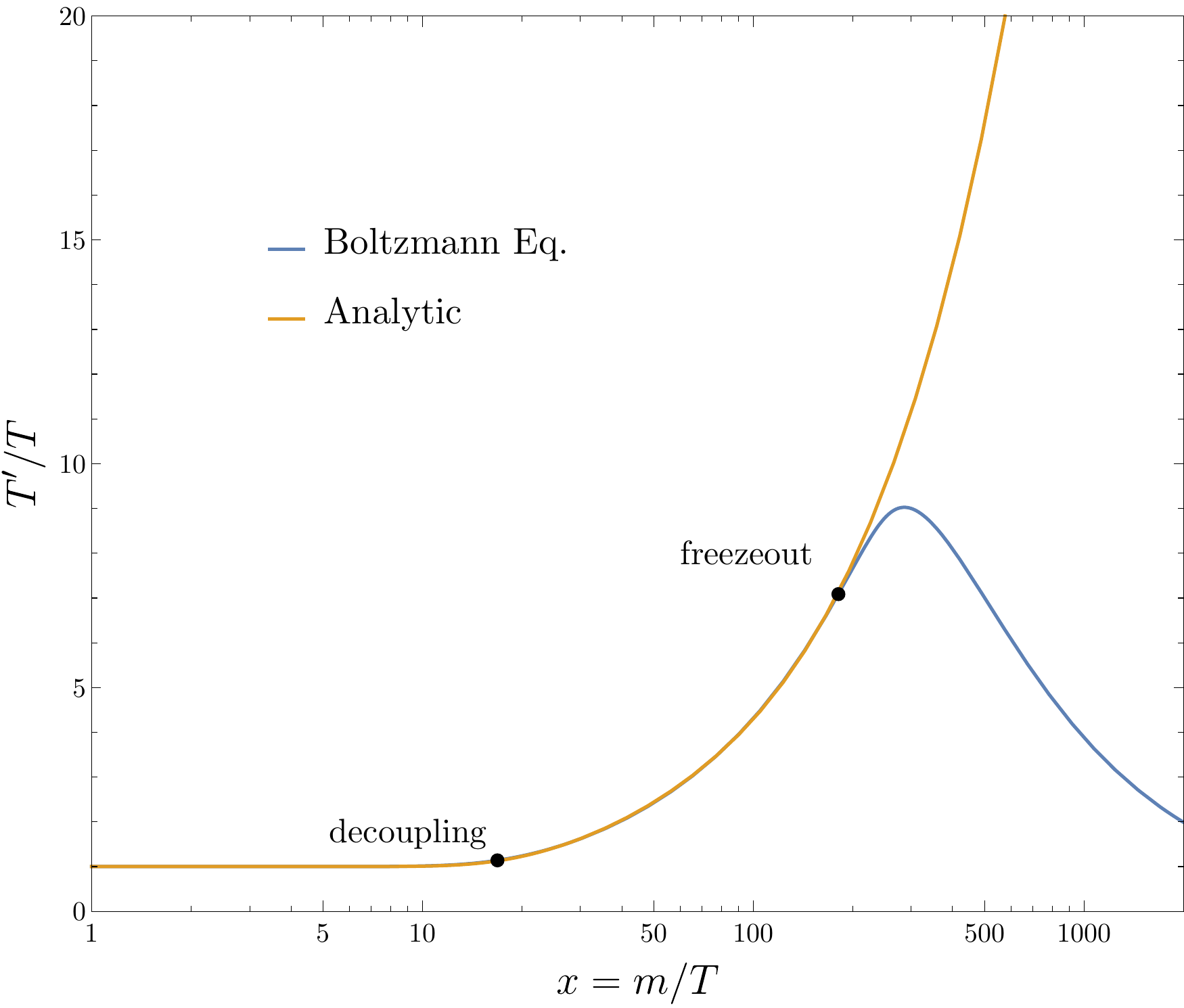}
\vspace{-2mm}
\caption{Evolution of the ratio of ELDER temperature $T^\prime$ to the SM plasma temperature $T$. Here $m_\chi=10$ MeV,  $c_2= 1.3\times 10^{-14}$, $\alpha = 5$, $g_\chi=2$, and $g_\psi=4$.}
\label{fig:evolution}
\end{center}
\end{figure}

The evolution of ELDER temperature throughout the kinetic decoupling and freeze-out process is illustrated in Fig.~\ref{fig:evolution}. The ELDER-to-SM temperature ratio starts growing after kinetic decoupling due to cannibalization, reaching the maximum value of $T^\prime/T\sim 10$ at the time of freeze-out. It drops rapidly after freeze-out since ELDERs are non-relativistic and $T^\prime \propto R^{-2}$, while $T\propto R^{-1}$, where $R$ is the size of the universe. The analytic function~\leqn{ANAL} provides an excellent approximation to the numerical solution of the Boltzmann equations up until $3\to 2$ freezeout.  

We note that Eq.~\leqn{cannib} can also be derived by assuming instantaneous kinetic decoupling between the dark sector and the SM at temperature $T_d$, and using the conservation of comoving entropy in the dark sector after decoupling. This approach was taken, for example, in Ref.~\cite{Kuflik:2015isi}. The alternative derivation presented here does not make the assumption of instantaneous decoupling, relying instead on the approximate solution for the evolution of $T^\prime$ accurate throughout the decoupling process. Apart from being better justified physically, the distinct advantage of the new derivation is that it automatically provides the expression for $T_d$ in terms of the underlying model parameters, Eq.~\leqn{Td}.  

In the instantaneous freeze-out approximation, the asymptotic value of the yield 
$Y_\chi=n_\chi/s_0$, where $s_0$ is the entropy density today, is given by
\beq
Y_\infty =  Y_{x_f }= \frac{g_\chi (2\pi x^\prime_f)^{-3/2} e^{-x^\prime_f}}{(2\pi^2/45) g_{* s, f} x_f^{-3}},
\eeq{yield} 
where $x_f$ and $x_f^\prime$ are the temperatures of the SM and the ELDERs, respectively, at the time of freeze-out. The effective multiplicity at freeze-out, $g_{*s, f}$, is strongly dominated by the SM degrees of freedom that are relativistic at that temperature, and the ELDER contribution to entropy is negligible; for typical ELDER parameters, $g_{*s,f}=10.75$. The ELDER relic density is given by
\beq
\Omega_\chi h^2 \simeq 3 \times 10^6 \, \left( \frac{m_\chi}{10~{\rm MeV}}\right)\, Y_\infty \simeq 
4 \times 10^5 \, \left( \frac{m_\chi}{10~{\rm MeV}}\right)\,  \frac{g_{\chi}}{g_{* s,\, f}} \frac{x_d^{3/2} e^{-x_d} }{\left(1+\frac{3}{x_d}\log \frac{x_f}{x_d}\right)^{3/2}}, 
\eeq{Omega}
where $x_d$ is the decoupling temperature defined in Eq.~\leqn{Td}. 

The $3\to2$ self-annihilations freeze-out when $n_\chi^2 \left< \sigma_{3\to 2} v^2 \right> \simeq H$. Let us parametrize 
\beq
\left< \sigma_{3\to 2} v^2 \right> \equiv \frac{\alpha^3}{m_\chi^5}.
\eeq{alpha}
The freeze-out  and decoupling temperatures can then be estimated by solving the equations
\beq
x_f^\prime + \frac{9}{4} \log x_f^\prime \simeq 31.0 - \frac{x_d}{2} - \frac{3}{4}\log \left( \frac{m_\chi}{10~{\rm MeV}}\right) + \frac{9}{4}\log\alpha + \frac{3}{2}(\log x_d + \log g_\chi-\frac{1}{4}\log \frac{g_{*,f}}{10})
\eeq{xfo}
and
\beq
x_d-3\log x_d \simeq 12.9 -\frac{3}{2}\log x_f^\prime + \log g_\chi - \log \frac{g_{*s,f}}{10}+ \log \left( \frac{m_\chi}{10~{\rm MeV}}\right) - \log \left( \frac{\Omega_\chi h^2}{0.1}\right).
\eeq{correct_xd}
Numerically, $x_d\simeq 17$ and $x_f^\prime\simeq 25$ for a typical ELDER model. The decoupling temperature is directly related to the strength of elastic scattering between ELDERs and SM particles, see Eqs.~\leqn{adef},~\leqn{Td}. Once $x_d$ is found by solving Eq.~\leqn{correct_xd}, it is straightforward to compute the corresponding elastic scattering strength: 
\beq
\bar{c}_n \simeq (1.4\times 10^{-18})\,\frac{g_{*,d}^{1/2} \xi_n}{g_\chi g_\psi^2}\, \left( \frac{m_\chi}{10~{\rm MeV}}\right)\,x_d^{n+4},
\eeq{correct_cn}
where $g_{*,d}$ is the effective number of relativistic degrees of freedom at $T_d$, and  
$
\xi_n=(n+4)[\Gamma(\frac{n+3}{n+4})]^{-n-4} /N^\psi_{3+n} 
$
is a numerical constant. (For future reference, $\xi_0 \simeq 0.08$ and $\xi_2\simeq 0.004$.) Once a mechanism that mediates ELDER-SM scattering is specified, this formula can be used to make detailed, robust phenomenological predictions, as discussed in the next Section. Remarkably, such predictions are almost completely independent of the details of  self-interactions of ELDERs, or their interactions with other dark sector states. 

\subsection{ELDER Mass Estimates}

A model-independent upper bound on the ELDER dark matter particle mass can be obtained as follows. 
Self-consistency of the ELDER scenario requires $x_f > x_d$, or\footnote{Close to this bound, the kinetic decoupling and freeze-out occur close in time, and the formulas derived in this Section, which assumed a clear separation between the two events, are not strictly applicable. The bound on $\alpha$ for ``pure ELDER" regime, in which the separation is clear, is stronger by a about a factor of 2. For smaller $\alpha$, a ``mixed SIMP-ELDER" regime occurs, which does not lend itself to simple analytic estimates. Numerical analysis of this regime indicates a smooth connection between ``pure SIMPs" and ``pure ELDERs", see for example Fig.~\ref{fig:regimes_old}.}
\beq
\alpha\gsim  0.5 \frac{m_\chi}{10~{\rm MeV}}.
\eeq{alpha_bound}
Here we see that ELDER dark matter is pushed to the strongly interacting regime ($\alpha \gtrsim 1$).
The thermally averaged $3\to 2$ rate can be bounded above by unitarity, in similar spirit to the bound derived on the thermally averaged WIMP annihilation rate~\cite{Griest:1989wd}. The optical theorem states that 
\beq
2~{\rm Im}\,\mathcal{M}_{\rm forward}  =  \sum_X \int  d\Pi_X (2\pi)^4 \delta^{4} (p_i -p_X) |\mathcal{M}_{\chi \chi \to X}|^2,
\eeq{optical_theorem}
where $\mathcal{M}_{\rm forward}$ is the matrix element for forward scattering $\chi \chi \to \chi \chi$, and $d\Pi_X$ is the Lorentz invariant phase space. (We assume that $\chi$ is a real scalar, and $d\Pi_X$ includes the relevant identical-particle factor.) Picking the term with $X= \chi \chi \chi $ from the sum yields the inequality 
\beq
\int  d\Pi_X (2\pi)^4 \delta^{4} (p_i -p_f) |\mathcal{M}_{3\to2}|^2 < 2~{\rm Im}\,\mathcal{M}_{\rm forward}. 
\eeq{2to3bound}
Using this in the definition of the thermally averaged rate in Eq.~\leqn{32thermavg}, in the non-relativistic limit, the rate is bounded above by
\beq
\left< \sigma_{3 \to2 } v^2\right> \lesssim  \frac{\sqrt{15}\pi}{ 12 T^3 m^{4}} e^{3m/T}\int_{9m^2}^\infty ds \,e^{-\frac{\sqrt{s}}{T}}\,{\rm Im}\,(\mathcal{M}_{\rm forward}(s)).
\eeq{unitaritybound1}
 In the absence of light degrees of freedom, non-relativistic elastic scattering of scalar $\chi$ particles is typically dominated by the $s$ wave. Partial-wave unitarity requires\footnote{At $\sqrt{s}=3m$, the $\chi$ particles are moderately relativistic, $\beta^2\sim0.5$, and corrections to $s$-wave scattering amplitude may be non-negligible. This will affect the unitarity bound at the level of order-one factors. Thus, this bound as well as the mass bound in Eq.~\leqn{mupper} should be viewed as order-of-magnitude estimates.} $|{\cal M}_{\rm forward}| \le 16 \pi \sqrt{s}/p \simeq 96 \pi /\sqrt{5}$, which in turn implies (taking into account the typical freeze-out temperature $x_f^\prime\simeq 20$) an upper bound 
\beq
\alpha \lsim 73.  
\eeq{unitarity}
where $\alpha$ is defined in Eq.~\leqn{alpha}. Combining this bound with Eq.~\leqn{alpha_bound} yields
\beq
m_\chi \lsim 1~{\rm GeV}.
\eeq{mupper}
This partial-wave unitarity bound is independent of the details of the dark sector. In specific models of dark sector self-interactions, other considerations, such as perturbativity of couplings, may impose stronger bounds. For example, in simple scalar models discussed in Section~\ref{sec:scalars}, the upper bound on the ELDER mass from perturbativity is about 200 MeV.

There is also a lower bound on $m_\chi$. As the ELDER becomes non-relativistic, energy and entropy are transferred from the dark sector to the SM, reheating the SM degrees of freedom. This process continues until the decoupling of elastic scattering between ELDERs and the SM at temperature $T_d$. If the energy and entropy transfer is active during or after Big-Bang Nucleosynthesis (BBN), it will generally result in modification of BBN predictions for light-element abundances, and/or the effective number of neutrinos $N_{\rm eff}$ inferred from the Cosmic Microwave Background (CMB) measurements; see {\it e.g.} Ref.~\cite{Boehm:2013jpa}. This is certainly the case if the interactions between the ELDER and the SM are mediated via the dark photon portal, which, as argued in Section~\ref{sec:pheno}, is the most plausible renormalizable portal compatible with this scenario. The dark photon portal couples the ELDERs very weakly to neutrinos. If entropy transfer continues below the temperature of neutrino decoupling from the electron/photon plasma, non-standard $N_{\rm eff}$ is produced. It is in principle possible that this bound could be avoided in a model in which electrons, photons and neutrinos are reheated equally. However in this paper we will adopt~\cite{Boehm:2013jpa}
\beq
m_\chi \gsim 5~{\rm MeV}
\eeq{mlower} 
as a rough lower bound on the ELDER mass.

\subsection{ELDERs, SIMPs and WIMPs, Oh My!}

If in a given model $c_n<\bar{c}_n$, defined in \leqn{cndef} and \leqn{correct_cn}, the particle $\chi$ cannot account for the observed dark matter. On the other hand, if $c_n>\bar{c}_n$, the correct relic density can still be achieved through the SIMP mechanism. In this case, dark matter and SM remain in kinetic equilibrium until the $3\to 2$ interactions decouple and the $\chi$ density freezes out: $x_d>x_f$. The relic density is given by
\beq
\Omega_\chi h^2 \simeq 0.02 \, \left( \frac{m_\chi}{10~{\rm MeV}}\right)^{3/2} \, \alpha^{-3/2}\,\left( \frac{x_f}{20}\right)^2,
\eeq{SIMP}
where the freeze-out temperature $x_f$ is found as a solution to
\beq
x_f+\frac{1}{2}\log x_f= 20.7 - \frac{1}{2}\log \left( \frac{m_\chi}{10~{\rm MeV}}\right) + \frac{3}{2} \log\alpha + \log g_\chi.
\eeq{xfo_SIMP}
After freeze-out, elastic scattering with SM no longer affects $n_\chi$; thus in the SIMP regime, the relic density is determined by the self-interaction strength $\alpha$, and is independent of $c_n$. The SIMP value of $\alpha$,
\beq
\alpha_{\rm SIMP} \simeq 0.34 \, \left( \frac{m_\chi}{10~{\rm MeV}}\right)\, \left( \frac{\Omega_\chi h^2}{0.1}\right),
\eeq{alpha_SIMP}
is close to the lower bound on $\alpha$ required for the ELDER scenario, Eq.~\leqn{alpha_bound}, and scales the same way with $m_\chi$. This gives a clear intuitive picture of the relation between the two regimes: for a given dark matter particle mass, the ELDER value of $c_n$ gives the lower bound on $c_n$ for SIMPs, while $\alpha_{\rm SIMP}$ is the lower bound of $\alpha$ for ELDERs. 

\begin{figure}[t!]
\begin{center}
\includegraphics[width=8.7cm]{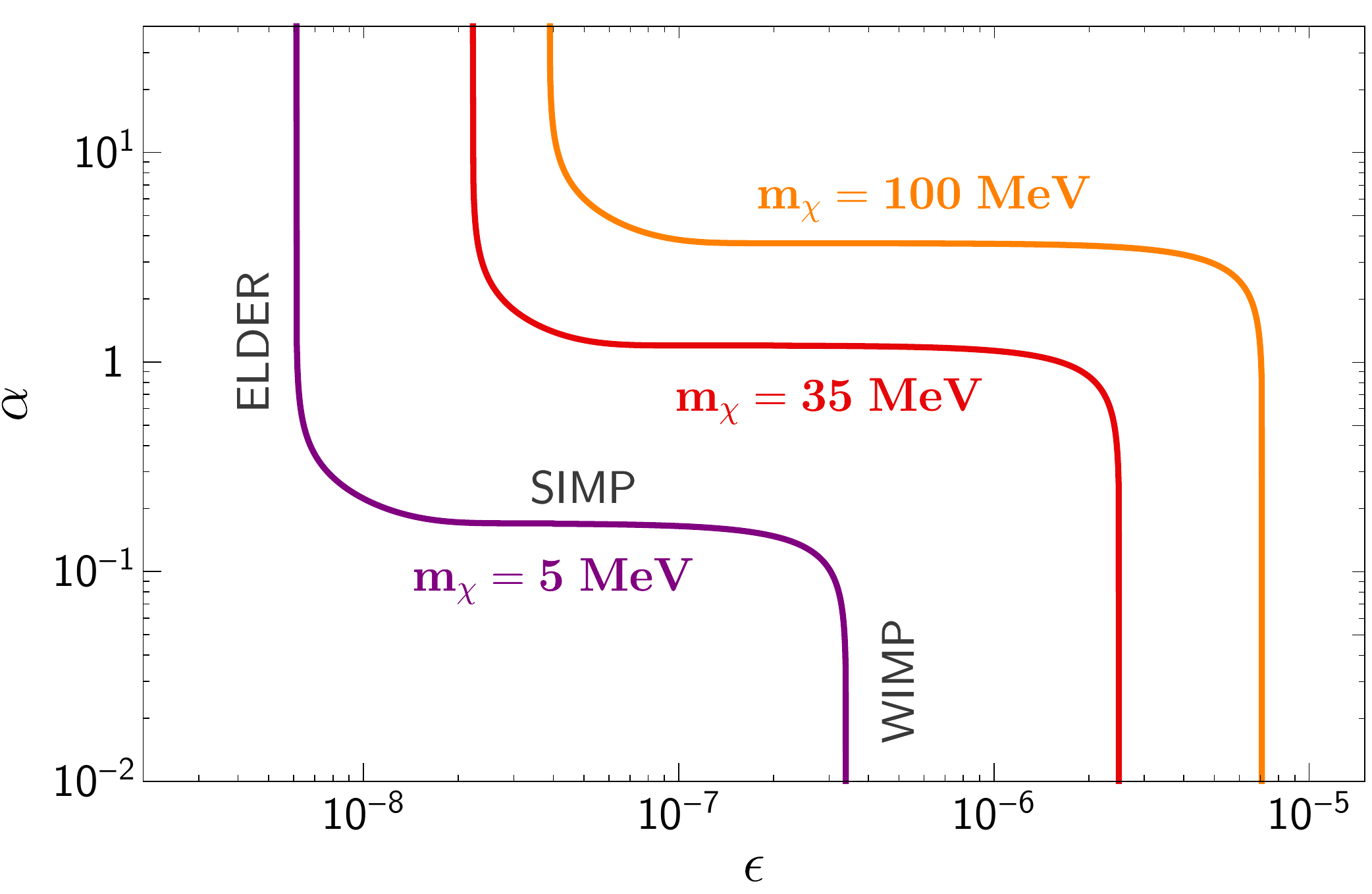}
\vspace{-2mm}
\caption{Regions of parameters corresponding to the observed relic density. For each mass, the vertical section of the line of the left/top corresponds to the elastically decoupling relic (ELDER) scenario proposed in this paper; the horizontal line to the SIMP scenario; and the vertical section on the right/bottom to the WIMP scenario. This figure, reproduced from Ref.~\cite{Kuflik:2015isi}, is based on a numerical solution of the Boltzmann equations for a model with $g_\chi=2$, $\psi=$ photon, $n=0$, $c_0= 8\pi\epsilon^2$. The same behavior is observed in other models, see for example Fig.~\ref{fig:Omega} below.  }
\label{fig:regimes_old}
\end{center}
\end{figure}

If $c_n$ is increased even further, eventually a point is reached where annihilations to SM decouple after the $3\to 2$ interactions. At this point, the relic density is determined by the cross section of annihilations to SM, and is once again independent of $\alpha$. Since this is the mechanism that sets the relic abundance of the conventional WIMPs, we refer to it as  the ``WIMP regime", even though the dark matter particle mass is still well below the weak scale, and a small coupling to SM is required to obtain the correct relic density. (For theoretically motivated realizations of such a scenario, see~\cite{Feng:2008ya}.) Figure~\ref{fig:regimes_old} illustrates the three regimes. This figure, reproduced from Ref.~\cite{Kuflik:2015isi}, is based on a numerical solution of the Boltzmann equations for a model with $g_\chi=2$, $\psi=$ photon, $n=0$, $c_0= 8\pi\epsilon^2$, which was performed in that paper. The same behavior is observed in other models, see for example Fig.~\ref{fig:Omega} below.  

\section{Dark Photon Portal and Phenomenology}
\label{sec:pheno}

It is well known that there are only three renormalizable interactions that can couple SM to dark sector states: ``dark photon", ``Higgs", and ``right-handed neutrino" portals~\cite{Alexander:2016aln}. Of these, only the dark photon portal is compatible with the ELDER scenario in its simplest form. In the case of the Higgs portal, the interaction has the form $S^2H^2$, where $S$ is a dark-sector field and $H$ is the SM Higgs. In the case of ELDER, the decoupling temperature is at the MeV scale, and the relevant SM degrees of freedom are electrons, photons, and neutrinos. The couplings to these particles at MeV temperatures mediated by the Higgs are too weak to produce the elastic scattering of the strength required in the ELDER scenario. In the case of the neutrino portal, the interaction is of the form $HLN$, where $N$ is a dark-sector fermion. The ELDER dark matter particle must possess $3\to 2$ interactions, and thus must be a boson. If the dark matter is a fermion, then cannibalization may occur via $4\to 2$ annihilations. However, this leads to strongly self-interacting sub-MeV DM, which is excluded by BBN and structure formation~\cite{Hochberg:2014dra}.  Hence, we will focus on the dark photon portal as the most plausible mechanism for ELDER-SM coupling. 

\subsection{Dark Photon Portal}

Specifically, we consider a complex scalar field $\chi$, neutral under SM gauge symmetries but charged under an abelian $U(1)_D$ gauge group in the dark sector:
\beq
{\cal L} = \left| D_\mu \chi \right|^2 = \partial^\mu\chi \partial_\mu\chi^* + i g_D A^\prime_\mu \left(\chi^* \partial_\mu \chi - \chi \partial_\mu \chi^*\right) + \ldots
\eeq{Lchi}  
where $g_D$ is the $U(1)_D$ coupling constant, and $A^\prime$ is the corresponding gauge field. The $A^\prime$ kinetically mixes with the SM photon:\footnote{In the fundamental theory, the mixing involves the SM hypercharge gauge field. Since the physics considered here takes place well below the weak scale, we ignore the mixing with the $Z$ boson.}
\beq
{\cal L}_{\rm k-m} = \frac{1}{2}\frac{\eg}{\cos\theta_W} B^{\mu\nu} F_{D\mu\nu},
\eeq{kinmix} 
where $B$ and $F_D$ are the field strength tensors of the $U(1)_Y$ and $U(1)_D$, and $\theta_W$ is the Weinberg angle. Diagonalizing the kinetic terms yields the SM photon $A$, under which $\chi$ is uncharged, and the ``dark photon" $V$, which couples to the SM electromagnetic current with strength $\eg e$, and to the ``dark" $U(1)_D$ current with strength $g_D$. We further assume that $U(1)_D$ is broken, giving the dark photon mass $m_V$. (For specific models that realize this setup, including ELDER self-interactions, see Section~\ref{sec:scalars}.) If dark photons have a significant abundance in the early universe at the time of ELDER decoupling and freeze-out, the physics of these processes becomes considerably more complicated: for example, co-annihilation processes may play an important role in transferring energy between the SM and the dark sector. To avoid these complications, we focus our attention on the ``pure ELDER" case, when the dark photon is significantly heavier than the dark matter particle. For concreteness, we assume $m_V>2m_\chi$. 

Elastic scattering of ELDER on electrons is mediated by the $t$-channel dark photon exchange. In the language of Section~\ref{sec:ELDER}, the dark photon portal model corresponds to $\psi=e^\pm$, $g_\psi=4$, $n=2$, and the elastic scattering strength is given by\footnote{Near the upper boundary of the ELDER mass region, $m_\chi\sim 1$ GeV, scattering off charged pions and muons is relevant during decoupling, and the formulas in this Section are modified by ${\cal O}(1)$ factors to include their contributions.}  
\beq
c_2 = \frac{2 e^2 \epsilon_\gamma^2 g_D^2 m_\chi^4}{m_V^4} \simeq 2.3 y.
\eeq{c2DP}  
Here we defined the dimensionless combination 
\beq
y=\epsilon_\gamma^2 \alpha_D \left(\frac{m_\chi}{m_V}\right)^4,
\eeq{y_def}
where $\alpha_D=g_D^2/(4\pi)$. This is the same combination of parameters that controls dark matter annihilations to the SM, as has been previously noticed in studies of the conventional scenario where such annihilations determine the relic density~\cite{Izaguirre:2015yja}. In the ELDER scenario, the value of $y$ that corresponds to the observed relic density can be inferred from Eq.~\leqn{correct_cn}: 
\beq
y_{\rm ELDER} \simeq 5.8 \times 10^{-15}\,\left( \frac{g_{*,d}}{10}\right)^{1/2}\,  \left( \frac{m_\chi}{10~{\rm MeV}}\right)\, \left( \frac{x_d}{17}\right)^6,
\eeq{yELDER} 
where $x_d$ is the solution to Eq.~\leqn{correct_xd}. This is a robust prediction of the ELDER scenario with the dark photon portal, independent of the details of ELDER self-interaction dynamics. 

As discussed above, if the dark matter coupling to the SM is increased above the ELDER value, correct relic density can still be achieved by SIMP or WIMP mechanisms. In the dark photon portal model, the WIMP regime corresponds to the well-known ``thermal target" value for $y$~\cite{Izaguirre:2015yja}: 
\beq
\frac{y_{\rm WIMP}}{\xi^2} \simeq 1.4 \times 10^{-11} \, \left( \frac{\Omega_\chi h^2}{0.1} \right)^{-1} \left( \frac{m_\chi}{10~{\rm MeV}}\right)^2  \left( \frac{x_{f,a}}{20}\right) ^2,
\eeq{yWIMP}
where $\xi = 1-4m_\chi^2/m_V^2$, and $x_{f,a}$ is the temperature at which annihilations to SM freeze out. 
Any value of $y$ between $y_{\rm ELDER}$ and the thermal target is compatible with the SIMP mechanism, which can yield the correct relic density for appropriately chosen $3\to 2$ self-scattering cross sections.

Before proceeding, let us briefly comment on the astrophysical constraints on this model. Dark matter pair annihilation into electrons is constrained by the CMB measurements~\cite{Finkbeiner:2011dx,Madhavacheril:2013cna,Slatyer:2015jla}, as well as indirect-detection searches. However, in the case of scalar dark matter in the relevant mass range, the $s$-wave annihilation cross section is suppressed by a factor of $(m_e/m_\chi)^2 \lsim 10^{-2}$, while the $p$-wave contribution is velocity-suppressed. As a result, ELDER dark matter is easily consistent with these constraints. Also, the reaction $e^+e^-\to \chi\chi$ (with or without an on-shell dark photon) can provide an additional mechanism of cooling in supernovae, which is constrained by the observation of neutrinos from SN1987A (see {\it e.g.}~\cite{Chang:2016ntp, Hardy:2016kme}). We checked that in the ELDER region, the elastic scattering of $\chi$ on electrons is always sufficiently strong to prevent the dark matter particles from leaving the supernova core. The produced $\chi$'s become trapped in the core, and do not contribute to the cooling rate.  

\subsection{Direct Detection}

\begin{figure}[t!]
\begin{center}
  \includegraphics[width=0.8\textwidth]{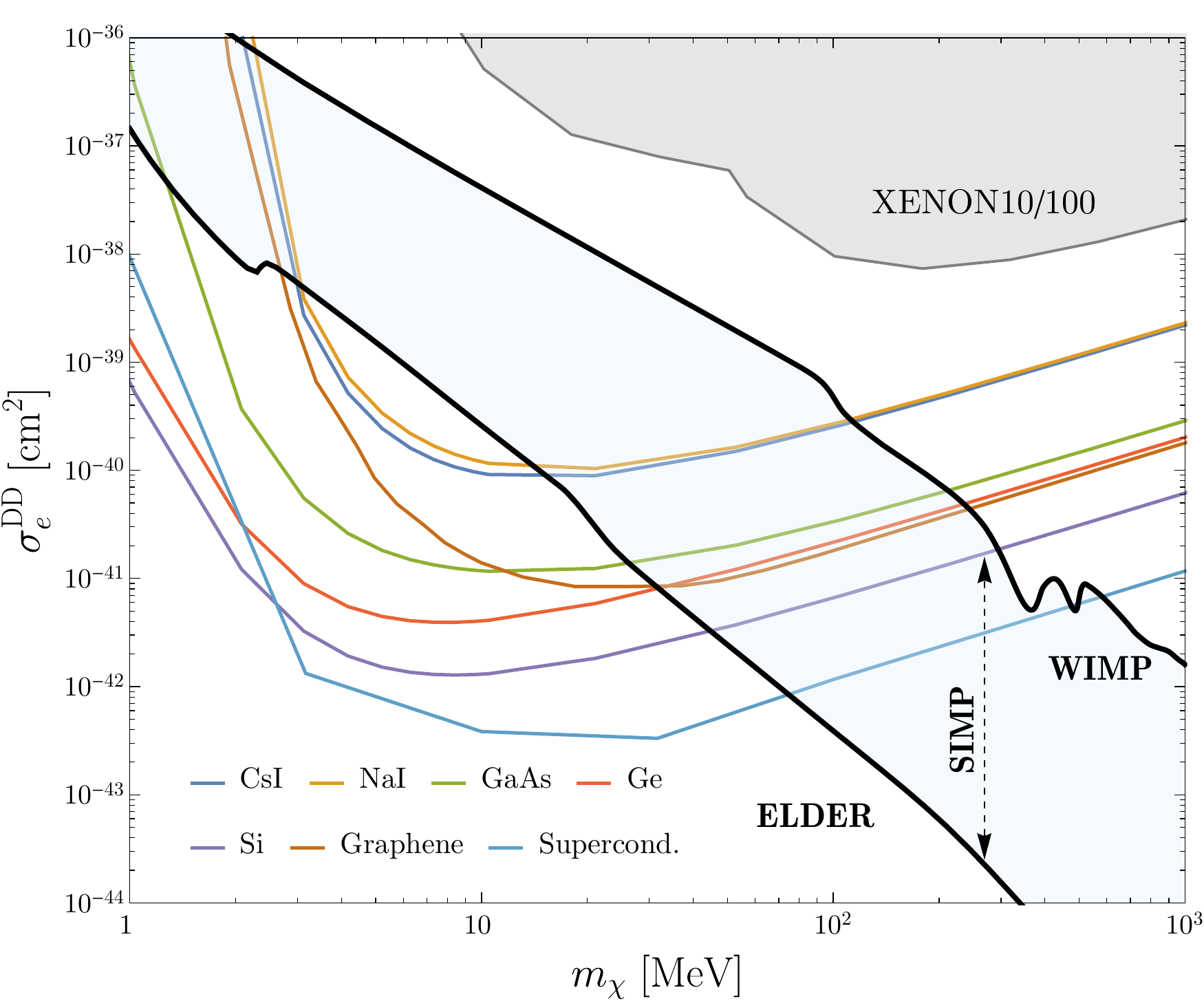}
  \end{center}
 \caption{Direct detection cross section, $\sigma^{\rm DD}_e $, predicted in the ELDER, WIMP and SIMP scenarios with a dark photon portal. For comparison, also shown are the current bounds from XENON experiment~\cite{Essig:2012yx,Essig:2017kqs} and projected sensitivities for 3 events in 1 kg-year exposure of proposed experiments: semiconductors~\cite{Essig:2011nj,Graham:2012su,Essig:2015cda,Tiffenberg:2017aac}, superconductors (10 meV threshold)~\cite{Hochberg:2015fth,Hochberg:2015pha}, superfluids~\cite{Schutz:2016tid,Knapen:2016cue}, scintillators~\cite{Essig:2011nj,Derenzo:2016fse} and graphene~\cite{Hochberg:2016ntt}.  \label{fig:DD}   
}
\end{figure}

Direct detection of sub-GeV dark matter has been an area of active recent investigations. Heavy nuclear recoils  do not carry sufficient energy to be detected in this mass range, and direct detection is easier for dark matter scattering on electrons. Remarkably, in the ELDER scenario with a dark photon portal, it is {\it precisely the same process} that determines the DM relic density. The observed dark matter density completely determines the direct detection cross section, with essentially no free parameters other than the ELDER mass $m_\chi$. The direct detection cross section is given by
\beq
\sigma^{\rm DD}_e = \frac{16 \pi \alpha m_e^2}{m_\chi^4}\,y.
\eeq{DD}
Setting $y=y_{\rm ELDER}$ in this formula defines a very sharp ``ELDER target" for the direct detection experiments. This complements the ``thermal target"~\cite{Izaguirre:2015yja,Alexander:2016aln}, which in our language corresponds to $y=y_{\rm WIMP}$, while the region $y_{\rm ELDER}<y<y_{\rm WIMP}$ corresponds to SIMP dark matter. Moreover, as discussed above, observational constraints and unitarity considerations restrict $m_\chi$ to a range between roughly 5 MeV and 1 GeV. These considerations define the {\it direct detection target region}, shown in Fig.~\ref{fig:DD}. 

The predicted cross sections are well below the current XENON bounds~\cite{Essig:2012yx,Essig:2017kqs}. However, novel experimental approaches that are currently being investigated have the potential to dramatically increase the sensitivity to DM-electron scattering in this mass range. Target materials under study include semiconductors~\cite{Essig:2011nj,Graham:2012su,Essig:2015cda,Tiffenberg:2017aac}, noble liquids~\cite{Essig:2011nj,Essig:2012yx}, superconductors~\cite{Hochberg:2015fth,Hochberg:2015pha}, superfluids~\cite{Schutz:2016tid,Knapen:2016cue}, scintillators~\cite{Essig:2011nj,Derenzo:2016fse} and graphene~\cite{Hochberg:2016ntt}. Projected sensitivities of these experiments will allow them to test a significant part of the SIMP and ELDER target region, see Fig.~\ref{fig:DD}. 

\subsection{Dark Photon Searches}

\begin{figure}[t!]
\begin{center}
  \includegraphics[width=0.8\textwidth]{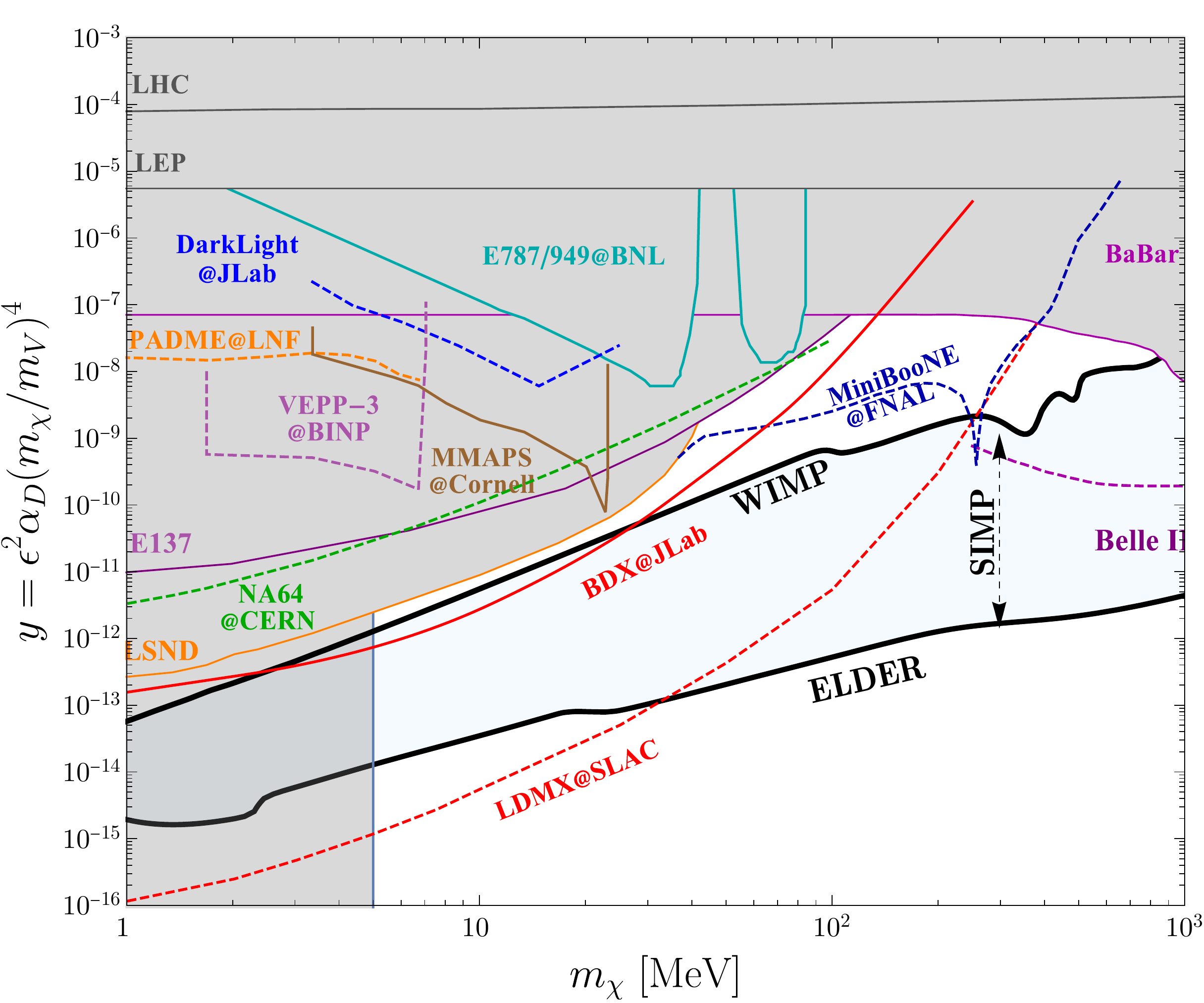}
  \end{center}
 \caption{The dark photon target region predicted in the ELDER, WIMP and SIMP scenarios. For comparison, the current bounds and projected sensitivities of searches for dark photon decaying to dark matter particles~\cite{Alexander:2016aln} are also shown.  \label{fig:DP}}
\end{figure}

Searches for a dark photon in the MeV-GeV range have also been an area of much activity recently. Existing experimental data has been used to place bounds on the dark photon, and several dedicated experiments are now running or in preparation. The ELDER, SIMP and WIMP scenarios with dark photon portal provide a well-defined {\it dark photon target region} for such experiments, shown in Fig.~\ref{fig:DP}. 

In the ELDER scenario, the dark photon mass $m_V$ must be large enough so that the process $\chi\chi^*\leftrightarrow VV$ is not relevant throughout the $\chi$ kinetic decoupling and freeze-out process. For the discussion of this section, we assume $m_V>2m_\chi$. In this case, the decay $V\to \chi\chi^*$ is likely to be the dominant dark photon decay channel, since its amplitude is proportional to the dark sector gauge coupling $g_D$, which is naturally of order one, while the amplitudes of competing decays such as $V\to e^+e^-$ are controlled by the small kinetic mixing parameter $\eg$. As a result, the experiments relevant for constraining our scenario are those searching for {\it invisible} dark photon decays. There are two basic experimental approaches. First, one can search for missing mass or energy in collider events due to an invisible particle $V$. The strongest current constraints from this approach come from re-analysis of BaBar data~\cite{Lees:2017lec}, as well as, at low masses, the dedicated NA-64 experiment at CERN~\cite{Banerjee:2016tad}. These searches do not yet constrain the ELDER scenario. In the future, the missing-energy LDMX experiment proposed at SLAC~\cite{Izaguirre:2015yja,Izaguirre:2014bca} will have sufficient sensitivity to test a significant part of the ELDER parameter space. Second, one can search for a dark matter particle that is produced in dark photon decay and propagates through shielding material to a downstream detector. (This would in effect amount to ``direct detection" of a dark matter particle produced in an accelerator.) This approach was recently pioneered by the MiniBooNE experiment~\cite{Aguilar-Arevalo:2017mqx}, and dedicated experiments such as BDX~\cite{Battaglieri:2014qoa} and SHiP~\cite{Alekhin:2015byh} have been proposed. Such future experiments may be sensitive to ELDER and SIMP dark matter. A snapshot of the current and expected sensitivities of a variety of dark photon searches, collected in Ref.~\cite{Alexander:2016aln}, and overlaid with the ELDER and SIMP regimes, is shown in Fig.~\ref{fig:DP}. 

We remind the reader that while the theoretical predictions of the dark photon target region are naturally defined in terms of the $y$ variable, and are largely insensitive to variations of model parameters that leave $y$ unchanged, the same is {\it not} true of experimental sensitivities, which depend on model parameters in different ways. For example, sensitivity of a missing-mass experiment such as BaBar is completely independent of $g_D$, as long as it's large enough so that the invisible branching ratio of $V$ is close to 100\%. Thus, additional assumptions have to be made in displaying experimental sensitivities in terms of $y$, as in Fig.~\ref{fig:DP}; see Ref.~\cite{Alexander:2016aln} for further discussion.


\section{Models of ELDERs}
\label{sec:scalars}

We argued above that strong self-interactions in the dark sector are required in the ELDER scenario, with $3\to 2$ cross-section of order one in its natural units. At some level, this is welcome: Such strong self-interactions are indeed expected if the ELDER is a bound state of confining dynamics in the dark sector, a paradigm that can potentially provide a natural explanation of proximity of the ELDER mass to the QCD confinement scale. On the other hand, it does create obvious challenges for model-building. Moreover, strong number-changing self-interactions tend to be accompanied by a large ELDER elastic scattering cross section, which can run afoul of observational constraints on dark matter self-scattering in galactic clusters such as the Bullet cluster. Fortunately, many phenomenological predictions of the ELDER scenario are independent of the details of dark sector self-interactions. This allowed us to completely sidestep these questions in the discussion of Section~\ref{sec:pheno}. We will now discuss two simple, renormalizable dark-sector models that explicitly realize the ELDER scenario. While not deeply rooted in strong gauge dynamics, they can be thought of as toy models representing interactions among the lightest mesons produced by such dynamics. They provide a useful illustration of the issues involved in dark-sector model building, and an ``existence proof" demonstration that consistent models can be found.  

\subsection{$\chi^3$ Model}

Here we consider a simple model in which the dark matter is a complex scalar charged under an unbroken $Z_3$ symmetry~\cite{Hochberg:2014dra,Choi:2015bya,Bernal:2015bla}.
Consider a dark sector consisting of a $U(1)_D$ gauge field with gauge coupling $g_D$, and two scalar fields charged under it, $\Phi$ and $\chi$, with $Q(\Phi)=+3$ and $Q(\chi)=+1$. The $\chi$ particle will play the role of dark matter. The scalar potential is 
\beq
V = V(\Phi) + V(\chi) + \frac{g}{3!} \left( \Phi^* \chi^3 + \Phi \chi^{*3}\right) + \lambda_{\Phi \chi} |\Phi|^2 |\chi|^2 ,
\eeq{Lsimple}
where $V(\psi)=m_\psi^2 |\psi|^2 + \lambda_\psi |\psi|^4$. We will assume $m_\Phi^2<0$, so that this field gets a vacuum expectation value (vev) $\left< \Phi \right> = w/\sqrt{2}$. We further assume that $m_\chi^2$ is positive. For simplicity, we consider the situation $m_\chi< |m_\Phi|$, with sufficient separation to ensure that the radial degree of freedom of $\Phi$ is sufficiently heavy to not play a role in the calculation of $\chi$ relic abundance. The effective Lagrangian for such calculation is then given by  
\beq
V_{\rm eff} = V(\chi) + \frac{R}{3!} m_\chi \left( \chi^3 +  \chi^{*3}\right),
\eeq{inter}
where we defined a dimensionless 3-point coupling
\beq
R = \frac{gw}{\sqrt{2}m_\chi}\,.
\eeq{R}
The only effect of the last term in the potential~\leqn{Lsimple} is  to renormalize the $\chi$ mass. 
The vev of $\Phi$ leaves a global $Z_3$ subgroup of the $U(1)_D$ unbroken, and the charge of $\chi$ under this discrete symmetry guarantees its stability, as required for a dark matter candidate. The $U(1)_D$ gauge boson gets a mass $m_V=\sqrt{3} g_D w$. The symmetry of the theory allows for kinetic mixing between the $U(1)_D$ gauge boson and the SM hypercharge gauge boson, as in Eq.~\leqn{kinmix}. As long as there are states, at any mass scale, that are charged under both gauge groups, such kinetic mixing will be generated, with values of $\epsilon_\gamma\sim 10^{-4}-10^{-2}$ being generic if no cancellations occur at the one-loop level~\cite{Holdom:1985ag}. Thus, this construction provides a stable scalar dark matter candidate with natural coupling to the electron via a dark photon portal. 

The matrix elements for non-relativistic $3\chi\to 2\chi$ annihilations are given by
\beq
{\cal M}(\chi\chi\chi^* \to \chi^*\chi^*) = -i \frac{13}{24} \frac{R^3}{m_\chi}\,,~~
{\cal M}(\chi\chi\chi \to \chi\chi^*) = +i \frac{1}{2} \frac{R^3}{m_\chi}\,.
\eeq{M32}
Here we set $\lambda_\chi=0$ for simplicity. This point is unexceptional (there is no enhanced symmetry associated with vanishing of $\lambda_\chi$) and is sufficient to illustrate the important physical features of the model. This yields the thermally averaged cross section
\beq
\left< \sigma v^2 \right> = \frac{\sqrt{5}}{2304\pi}\frac{265}{768} \frac{R^6}{m_\chi^5} \simeq 10^{-4} \frac{R^6}{m_\chi^5}.
\eeq{sv2}
In the SIMP scenario, the coupling $R$ can be inferred from the relic density as follows:
\beq
R_{\rm SIMP} \simeq 2.6\, \left(\frac{m_\chi}{10~{\rm MeV}}\right)^{1/2} \,\left(\frac{\Omega_\chi h^2}{0.1} \right)^{1/2}.
\eeq{Rsimp}
The required coupling is quite large, consistent with the idea that SIMP/ELDER dark matter particle can be a bound state of dark-sector confining gauge group: in this scenario, the potential~\leqn{Lsimple} can be thought of as a toy model representing the interactions among the two lightest mesons. The range of validity of the perturbative $\chi^3$ model can be estimated as $R\lsim4\pi$. In the SIMP scenario, this gives an upper bound on the dark matter particle mass:
\beq
m_\chi \lsim 230~{\rm MeV}. 
\eeq{chi3_mass_bound}
As discussed in Section~\ref{sec:ELDER}, the ``pure ELDER" scenario requires larger $3\to 2$ cross section than SIMP for the same $m_\chi$, and therefore the upper bound on $m_\chi$ is somewhat lower for ELDERs.  

The dark matter elastic self-scattering cross section is constrained by observations of galactic clusters, such as the Bullet cluster~\cite{Clowe:2003tk,Markevitch:2003at,Randall:2007ph}, and halo shapes~\cite{Rocha:2012jg,Zavala:2012us,Peter:2012jh}:
\beq
\frac{\bar{\sigma}_{2\to 2}}{m_\chi} \leq 0.47~{\rm cm}^2/{\rm g},
\eeq{sigmam_bound} 
where $\bar{\sigma}_{2\to 2}\equiv (\sigma(\chi\chi\to\chi\chi) + \sigma(\chi\chi^*\to\chi\chi^*))/2$. The $\chi^3$ model in the SIMP scenario predicts
\beq
\frac{\bar{\sigma}_{\rm SIMP}}{m_\chi} \,\simeq\, \left( \frac{m_\chi}{10~{\rm MeV}}\right)^{-1} \cdot \left( 30~\frac{{\rm cm}^2}{{\rm g}}\right),
\eeq{somSIMP}
while in the ELDER scenario the cross-section is even larger (bounded from below by Eq.~\leqn{somSIMP}). Thus, the simplest single-field $\chi^3$ model cannot provide sufficiently strong self-interactions required in these scenarios, while being consistent with observational constraints. We will now show that adding another dark-sector scalar field can resolve this problem. 

\subsection{Choi-Lee Model}
\label{sec:CL}

This model was originally introduced by Choi and Lee (CL)~\cite{Choi:2016hid} in the context of the SIMP scenario. The dark sector contains a $U(1)_D$ gauge symmetry, with gauge coupling $g_D$, and three complex scalar fields charged under this symmetry: $\phi$, $S$, and $\chi$, with charges $q_\phi=+5$, $q_S=+3$, and $q_\chi=+1$. The most general renormalizable scalar potential consistent with these charge assignments is
\beqa
V_{\rm d} &=& m_\phi^2|\Phi|^2 + \lambda_\phi |\Phi|^4+m_S^2|S|^2 + \lambda_S |S|^4 +m_\chi^2|\chi|^2 + \lambda_\chi |\chi|^4 \CR
& & + \lambda_{\phi S} |\Phi|^2|S|^2 + \lambda_{\phi\chi} |\Phi|^2|\chi|^2 + \lambda_{S\chi} |S|^2|\chi|^2 + \CR
& & + \frac{1}{\sqrt{2}} \lambda_1 \Phi^\dagger S^2 \chi^\dagger + \frac{1}{\sqrt{2}} \lambda_2 \Phi^\dagger S \chi^2 + \frac{1}{6} \lambda_3 S^\dagger \chi^3 + {\rm h.c.} 
\eeqa{V_high}     
We assume that $m_\phi^2<0$, while the other two scalar fields have positive mass-squared. The vev $\langle\Phi\rangle=w/\sqrt{2}$ breaks the gauge symmetry, giving the $U(1)_D$ gauge boson a mass $m_V=\sqrt{5} g_D w$. The $\Phi$ vev preserves a discrete $Z_5$ subgroup of the $U(1)_D$, under which $S$ and $\chi$ are both charged. The lighter of these particles, which we will assume to be the $\chi$, is therefore stable, and can play the role of dark matter. The scalar interactions after spontaneous symmetry breaking are described by
\beqa
V_{\rm d} &=& \frac{m_\chi}{\sqrt{2}} R_1 S^2 \chi^\dagger + \frac{m_\chi}{\sqrt{2}} R_2 S \chi^2 + \frac{1}{6} \lambda_3 S^\dagger \chi^3 + {\rm h.c.} \CR
& & + \lambda_S |S|^4 +  \lambda_\chi |\chi|^4 + \lambda_{S\chi} |S|^2|\chi|^2,
\eeqa{V_low} 
where we have omitted interactions with the Higgs component of $\Phi$ which play no role in the phenomenology considered here, and defined dimensionless couplings
\beq
R_i = \frac{v_D \lambda_i}{\sqrt{2} m_\chi},~~~i=1, 2.
\eeq{Rdef}
As in the $\chi^3$ model, the dark gauge boson $V$ kinetically mixes with the SM photon, providing a dark photon coupling between the dark sector and the SM. 

\begin{figure}
\begin{center}
\includegraphics[width=0.48\textwidth]{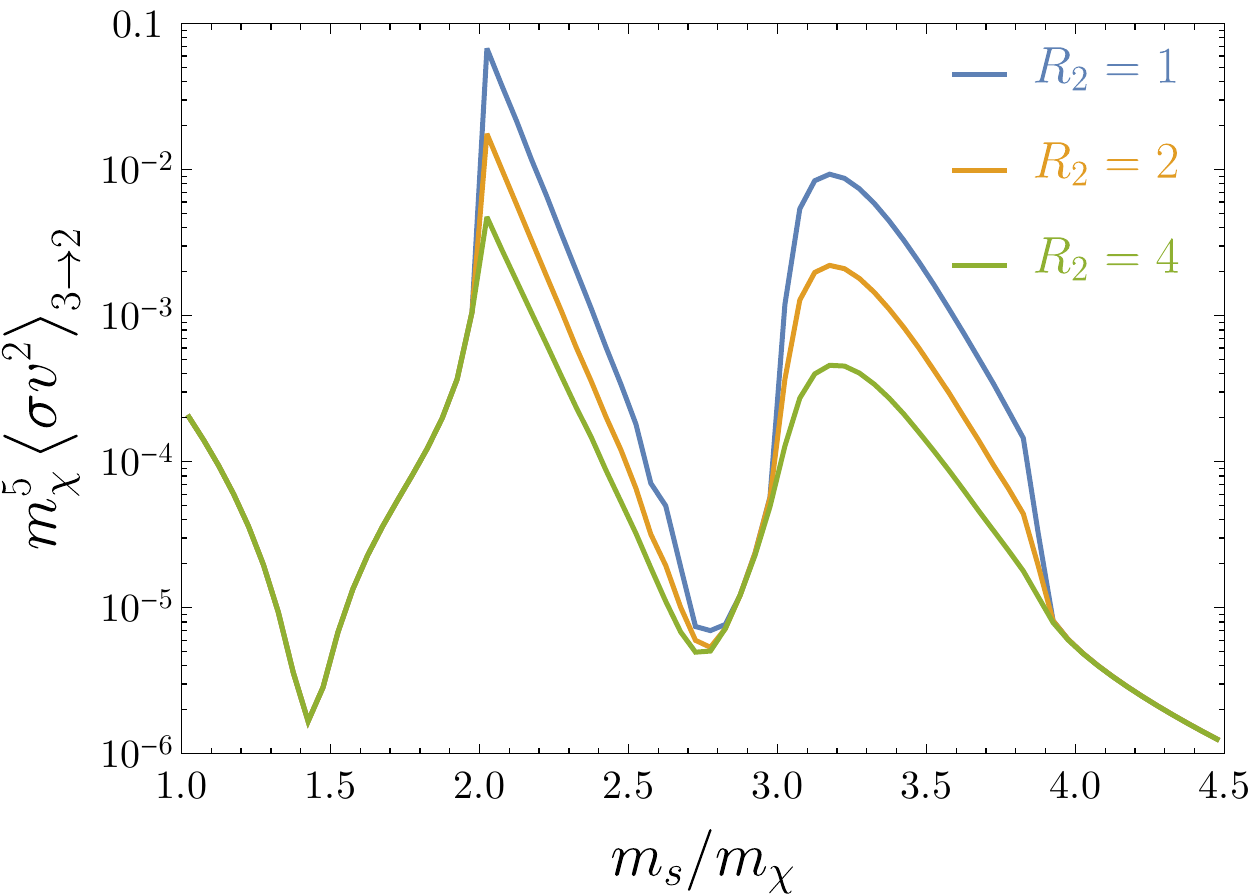}\hfill
\includegraphics[width=0.48\textwidth]{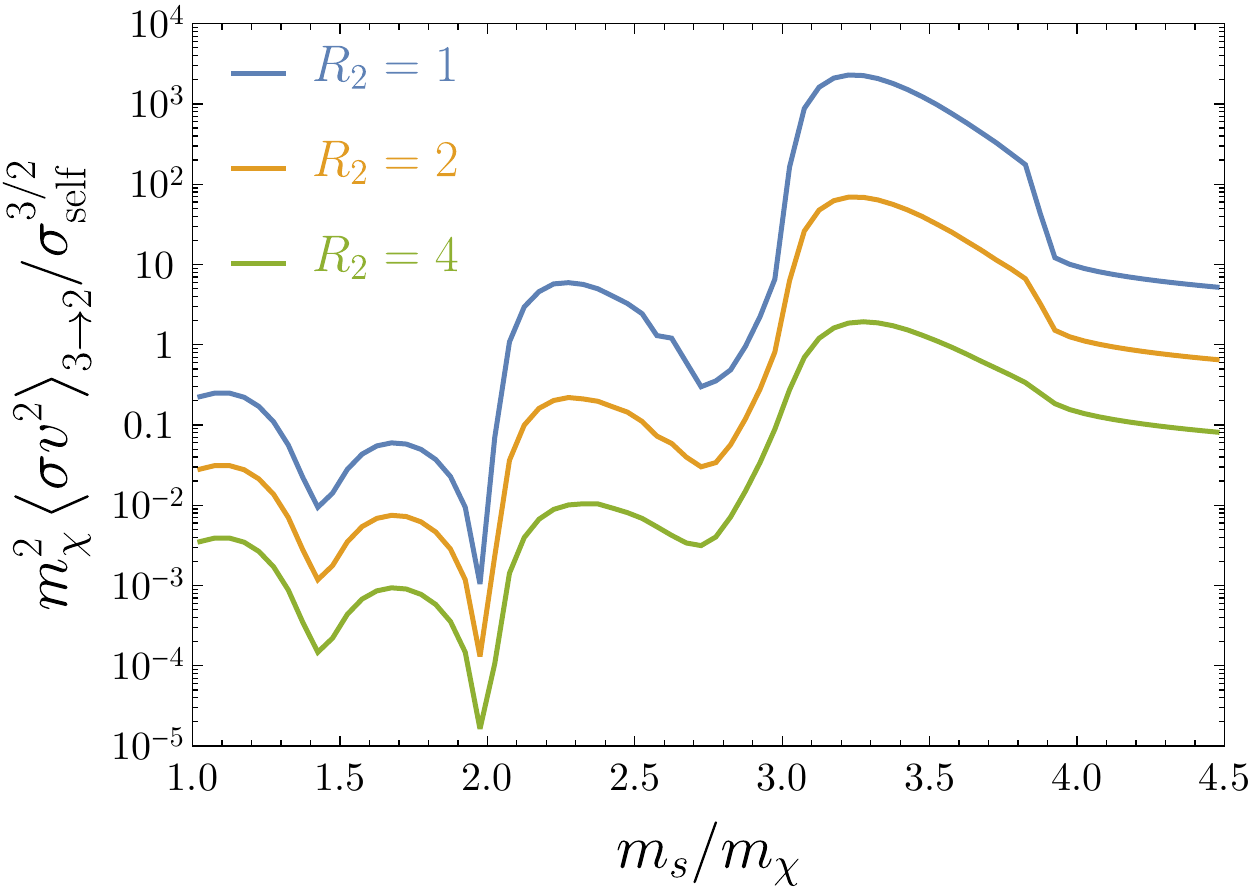}
  \end{center}
 \caption{The cross section $\langle \sigma_{3\to 2} v^2\rangle$ (left) and the dimensionless ratio of number-changing and elastic self-scattering rates (right), as a function of $m_S/m_\chi$. \label{fig:mSmchi}}
\end{figure}

The $2\chi\leftrightarrow 3\chi$ scattering process is induced by the couplings in the first line of Eq.~\leqn{V_low}. For simplicity, we set $\lambda_3=0$; this point is unexceptional (there is no enhanced symmetry associated with vanishing of $\lambda_3$) and is sufficient to illustrate the features of interest to us. The key observation is that for $m_S\approx 3m_\chi$, the $2\chi\leftrightarrow 3\chi$ scattering is resonantly enhanced, while the $2\chi\leftrightarrow 2\chi$ process is not. This effect is illustrated by the left panel of Fig.~\ref{fig:mSmchi}, where we plot the thermally-averaged $\langle \sigma_{3\to 2} v^2\rangle$ at temperature close to ELDER kinetic decoupling. A dimensionless ratio of the number-changing and number-preserving cross sections, $m_\chi^2 \langle \sigma_{3\to 2} v^2\rangle/\sigma_{2\to 2}^{3/2}$, can reach ${\cal O}(10^3)$. For comparison, in the $\chi^3$ model studied in the previous section, this ratio is close to 1. Note that the values of couplings $R_i$ required in the SIMP/ELDER scenarios are fairly large, so that the $S$ resonance is rather broad and no significant fine-tuning of $m_S/m_\chi$ is required to achieve significant enhancement of the $3\to 2 $ rate. This enhancement makes it possible to successfully implement SIMP and ELDER dark matter in the CL model without conflict with observational constraints from galaxy clusters and halo shapes.

Because of the resonance at $\sqrt{s} \approx 3m_\chi$, the quantity $\langle \sigma_{3\to 2} v^2\rangle$ has a non-trivial temperature dependence in the non-relativistic regime, making the parametrization of Eq.~\leqn{alpha} inapplicable. To compute the relic density, we integrate the Boltzmann equations numerically. 
The relic density is controlled by the seven model parameters that enter the Boltzmann equations: particle masses $m_\chi$, $m_S$, and $m_V$; and dimensionless coupling constants $R_1$, $R_2$, $g_D$, and $\eg$. To perform numerical analysis in this large parameter space, we made the following choices:

\begin{figure}[t!]
\begin{center}
\includegraphics[width=0.48\textwidth]{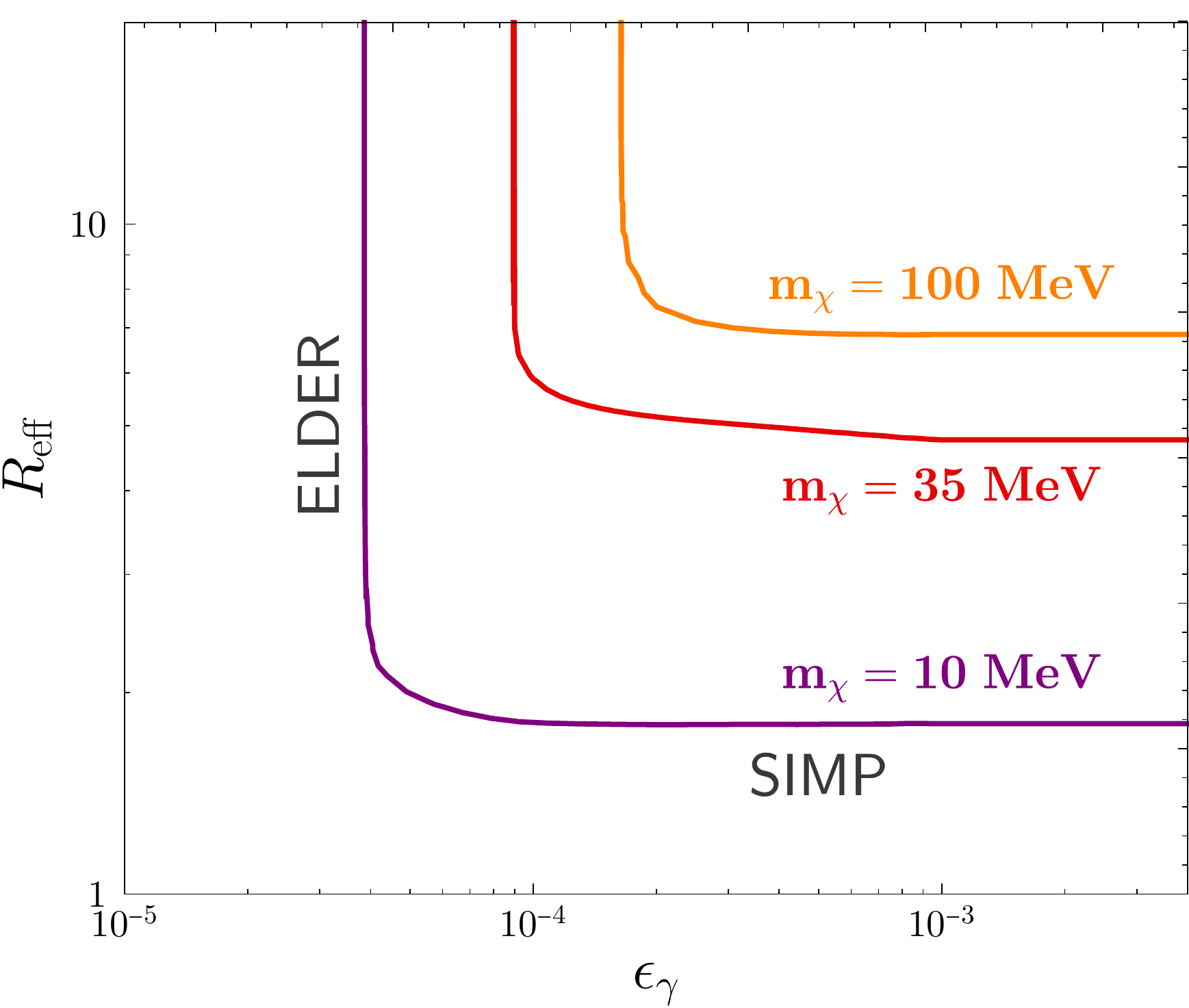}\hfill
  \includegraphics[width=0.471\textwidth]{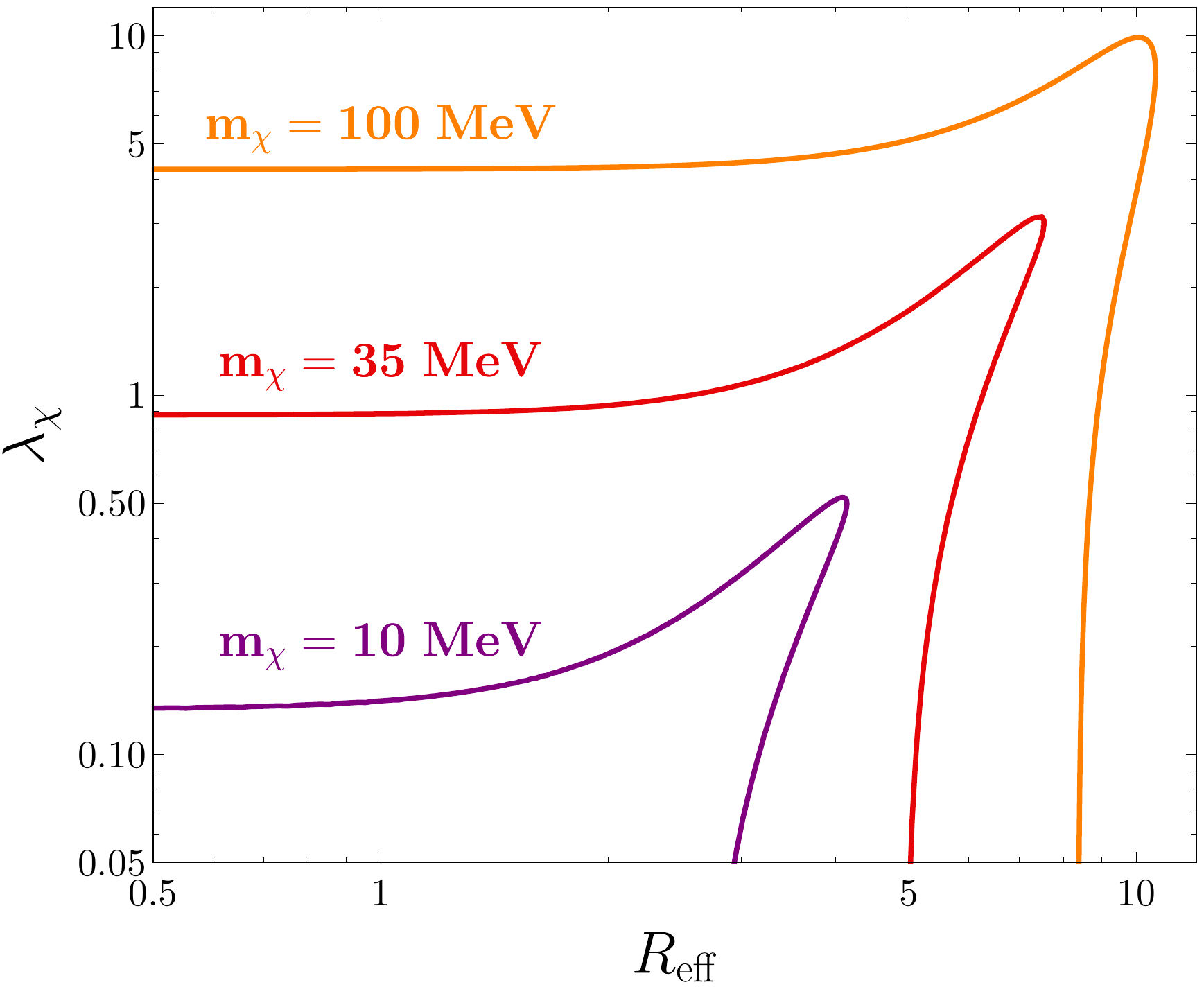}
  \end{center}
 \caption{Left: Regions of CL model parameter space with $\chi$ relic density consistent with the current best-fit $\Lambda$CDM value. Right: Constraints from galaxy cluster observations (the regions below the curves are allowed). In both plots, $m_S/m_\chi=3.1$, $m_V/m_\chi=10$, and $g_D=1$. In the right panel, we fix $R_1=10$ for illustration.
 \label{fig:Omega}}
\end{figure}

\begin{itemize}

\item The ratio of $S$ and $\chi$ masses was fixed close to the $3\to 2$ resonance, $m_S/m_\chi=3.1$. 

\item As discussed in Section~\ref{sec:pheno}, the relic density depends on the three parameters of the dark photon portal only through a single dimensionless combination $y$, defined in Eq.~\leqn{y_def}. Therefore it is sufficient to fix two of these parameters, and vary the third one. We fix $m_V/m_\chi=10$ and $g_D=1$, and vary $\eg$. 

\item The $3\to 2$ matrix element is proportional to a product $R_1R_2^2$, so that the DM relic density primarily depends on these couplings through the ``effective" $3\to 2$ coupling,
\beq
\Ref \equiv \left(R_1 R_2^2 \right)^{1/3}.
\eeq{Reff}
The relic density also depends on the width $\Gamma_S$, which is proportional to $R_2^2$. In practice, in the numerical analysis we fix $R_2$ (specifically, $R_2=2$ for $m_\chi=10$ MeV and $R_2=4$ for $m_\chi=35, 100$ MeV) and vary $R_1$. However, we checked that in the parameter range of interest, the relic density is insensitive to variations of $R_2$ within broad ranges around these values, allowing us to present the results solely in terms of the effective coupling $\Ref$.  

\end{itemize}

The three remaining parameters  $(m_\chi, \Ref, \eg)$, are scanned over. The results are illustrated in the left panel of Fig.~\ref{fig:Omega}, which shows regions of parameter space consistent with the current best-fit $\Lambda$CDM dark matter density,  $\Omega_\chi h^2=0.1188\pm0.0010$~\cite{Ade:2015xua}, in the $\eg-\Ref$ plane for three values of DM mass, 10, 35 and 100 MeV. The (roughly) horizontal bands of viable parameter space correspond to the SIMP scenario, while the (roughly) vertical bands realize the ELDER scenario. The values of $\epsilon_\gamma$ for the ELDER regime are in excellent agreement with the results of the analytic approach, Eq.~\leqn{yELDER}. In the intermediate regime, the DM-SM elastic scattering and the DM number-changing self-scattering decouple at roughly the same time, and both processes play a role in determining the relic density.  

As expected, realizing ELDER (or SIMP) dark matter in the CL model requires ${\cal O}(1)$ couplings among the scalars of the dark sector. The range of validity of perturbative CL model can be estimated as $R_1\lsim 4\pi$, $R_2\lsim 4\pi$. Combined with the relic density calculation, these constraints place an upper bound on the ELDER dark matter mass, $m_\chi \lsim 200$ MeV. Furthermore, the $\chi$ self-scattering cross section is constrained by observations of galactic clusters and halo shapes, Eq.~\leqn{sigmam_bound}. The self-scattering cross section receives contributions from a quartic coupling $\lambda_\chi$ as well as the $S$-exchange diagram controlled by $R_2$, and partial cancellation of the two diagrams is possible. Combined with the perturbativity bound on $R_1$, cluster observations place an upper bound on $\Ref$, shown in Fig.~\ref{fig:Omega}. For $m_\chi>5$ MeV, the values of $\Ref$ required in the ELDER scenario are compatible with observations.


\section{Conclusions}
\label{sec:conc}

In this paper, we studied the Elastically Decoupling Relic (ELDER) scenario for thermal dark matter. We presented an approximate analytic solution for the evolution of ELDER temperature throughout the kinetic decoupling epoch. This solution was used to provide explicit formulas relating various relevant quantities, such as, for example, the relic density of ELDERs and the cross section of their elastic scattering off SM particles. We also applied partial-wave unitarity constraint to obtain a bound on the allowed mass range for the ELDER dark matter candidate, 5 MeV $\lsim m_\chi \lsim$ 1 GeV. These results are valid in a broadly model-independent framework. 

Further, we showed that a dark photon portal can naturally provide the coupling between the dark matter particles and SM of the strength required in the ELDER scenario. Within the dark photon model, the ELDER scenario provides unambiguous predictions for dark matter direct detection experiments and dark photon searches, shown in Figs.~\ref{fig:DD} and~\ref{fig:DP}. These predictions have no free parameters other than the dark matter mass. They are also independent of the details of dark sector, as long as it provides sufficiently strong number-changing self-interactions to realize the ELDER scenario. Together with the well-known ``thermal target" and predictions of the Strongly-Interacting Massive Particle (SIMP) scenario, the ELDER predictions delineate a well-defined target region in the parameter spaces relevant for direct-detection and dark photon searches, which will be explored by the next generation of experiments. 

Both the ELDER and SIMP scenarios require ${\cal O}(1)$ strength (in natural units) of self-interactions among dark matter particles. Here, we studied two simple scalar-field models that incorporate such interactions while remaining within perturbative regime, for $m_\chi\lsim 200$ MeV. The models also naturally contain coupling to the SM via the dark photon portal. The simplest model, with just two scalar fields, exhibits tension with bounds on dark matter self-scattering cross section from observations of galaxy clusters. However, a slightly more complex model, with three scalar fields and a resonance structure, easily evades such bounds. These results indicate that there is no fundamental obstruction to finding dark sectors compatible with ELDER and/or SIMP scenarios.

 An important motivation for SIMP and ELDER scenarios is the proximity of the predicted dark matter particle mass to $\Lambda_{\rm QCD}$, a well-established important scale in the SM. In the toy models studied in this paper, $m_\chi\sim\Lambda_{\rm QCD}$ is put in by hand. The natural next step in the model-building direction would be to construct models in which this relation, as well as the strong self interactions, emerge naturally from UV physics. 

\section*{Acknowledgements}

We are grateful to Asher Berlin, Nikita Blinov, Rouven Essig, Yonit Hochberg, Hyun Min Lee, Tracy Slatyer, and Natalia Toro for useful discussions. This research is supported by the U.S. National Science Foundation through grant PHY-1316222, a grant \#399528 from the Simons Foundation (MP), and the Bethe Postdoctoral Fellowship (EK). YT is partially supported by the Visiting Graduate Fellow program at Perimeter Institute.

\appendix
\section{Boltzmann Equations}
\label{app:BEs}

The Boltzmann equation for the DM phase space distribution, $f_{\chi}({\bf p}; t)$, in an expanding Universe is  
\beq
\frac{\partial f_{\chi }}{\partial t} - H \frac{p^2}{E} \frac{\partial f_{\chi}}{\partial E} = C[f_{\chi}]\,,
\eeq{BEf}
where $E=\sqrt{{\bf p}^2+m_\chi^2}$, $H(t)$ is the Hubble expansion rate, and $C[f_{\chi}]$ is the collision term. For ELDER dark matter, the relevant collision terms are the $3 \to 2$ 
self-annihilations and $\chi$-$\psi$ elastic scattering ($\psi$ can be any light SM particle). The collision terms also includes annihilations to SM, but their effect in the ELDER scenario is negligible, and are omitted. Strong elastic self-scattering of ELDERs ensures that, throughout the kinetic decoupling and freeze-out process, the phase space distribution follows a thermal distribution:
\beq 
f_\chi = \frac{1}{e^{(E-\mu_\chi)/T^\prime} -1}\,,
\eeq{fchi}
where $\mu_\chi(t)$ is the chemical potential, and $T^\prime(t)$ is the temperature of the dark sector. Eq.~\leqn{BEf} can be most easily solved by taking the first two moments, the DM number density $n$ and energy density $\rho$:
\beq
 n_\chi = g_\chi \int \frac{d^3 p}{(2\pi)^3} f_\chi, ~~~~~~~~~  \rho_\chi = g_\chi \int \frac{d^3 p}{(2\pi)^3} E f_\chi,
\eeq{nandrho}
where $g_\chi$ is the number of degrees of freedom in $\chi$. These obey   
\beq
\frac{\partial n_{\chi}}{\partial t } + 3 H n_{\chi} = - \langle \sigma_{3 \to 2} v^2\rangle \left( n_{\chi}^3 - n_{\chi}^2 n_{\chi}^{\rm eq} \right), 
\eeq{BEn}
\beq
\frac{\partial \rho_{\chi}}{\partial t} + 3 H \left(\rho_{\chi} + P_{\chi}  \right) = -\langle \sigma_{el} v \delta E\rangle n_{\chi} n_{\rm \psi},
\eeq{BErho}
where $n_{\chi}^{\rm eq}$ is the density of $\chi$ in chemical equilibrium ({\it i.e.} at zero chemical potential). The thermally averaged $3\to2$  annihilation and energy transfer rates are
 \beqa
{n_{\chi}^{ 3}}  \langle \sigma_{3\to 2} v^2 \rangle  &=& \frac{1}{3! 2!} \int d\Pi_{\chi_1} d\Pi_{\chi_2} d\Pi_{\chi_3} d\Pi_{\chi_4} d\Pi_{\chi_5} (2\pi )^4\delta ^4\left(p_{\chi_1}+p_{\chi_2}+p_{\chi_3}-p_{\chi_4}-p_{\chi_5} \right) \nn\\
&&~~~\times f_{\chi_1} f_{\chi_2} f_{\chi_3}\overline{\left|\mathcal{M}_{\chi_1 \chi_2 \chi_3\to \chi_4 \chi_5}\right|^2}\,,
\eeqa{32thermavg}
 \beqa
{n_{\chi}  n_{\rm \psi} }\langle  \sigma_{el} v \delta E\rangle &=& \int d\Pi_{\chi_1} d\Pi_{\psi_1} d\Pi_{\chi_2} d\Pi_{\psi_2} (2\pi )^4\delta ^4\left(p_{\chi_1}+p_{\psi_1}-p_{\chi_2}-p_{\psi_2} \right)  \nn\\
&&~~~\times (E_{\chi_2} - E_{\chi_1}) f_{\chi_1} f_{\psi_1} \overline{\left|\mathcal{M}_{\chi_1 \psi_1 \to \chi_2 \psi_2}\right|^2} \,,
\eeqa{BEthermavg}
 where 
 \beq
 d \Pi_i \equiv \frac{g_i d^3 p_i}{(2\pi)^32 E_i}
 \eeq{dPi}
is the Lorentz invariant phase-space integration volume. The squared matrix elements, $\overline{\left|\mathcal{M}\right|^2}$, are averaged over initial and final degrees of freedom, including spin, color, and charge.\footnote{For the case of complex $\chi$ considered in this paper, we treat $\chi$ and $\chi^*$ as two states of the same particle, and averaging over these two states for each initial and final dark matter particle. For instance, for self-scattering
\beqa
\overline{|\mathcal{M}_{\chi_1\chi_2\to \chi_3\chi_4}|^2} &\equiv& \frac{1}{2^4} \Bigl( |\mathcal{M}(\chi_1\chi_2\to\chi_3\chi_4)|^2 + |\mathcal{M}(\chi_1^*\chi_2^*\to\chi_3^*\chi_4^*)|^2 + |\mathcal{M}(\chi_1\chi^*_2\to\chi_3\chi^*_4)|^2 \CR & &
\hskip-1cm + |\mathcal{M}(\chi_1\chi^*_2\to\chi^*_3\chi_4)|^2 + |\mathcal{M}(\chi^*_1\chi_2\to\chi_3\chi^*_4)|^2 +|\mathcal{M}(\chi_1\chi^*_2\to\chi^*_3\chi_4)|^2 \Bigr).
\eeqa{example}     
and for $\chi^{(*)}e^\pm \to \chi^{(*)}e^\pm$  in the dark photon portal,
\beq
\overline{\left|\mathcal{M}_{\chi_1 \psi_1 \to \chi_2 \psi_2}\right|^2} = \frac{1}{2^2}\frac{1}{4^2} \Bigl(|\mathcal{M}(\chi e^-)|^2 + |\mathcal{M}(\chi e^+)|^2+|\mathcal{M}(\chi^* e^-)|^2+|\mathcal{M}(\chi^* e^-)|^2 \Bigr)  =
\frac{e^2 g_D^2 \epsilon_\gamma^2m_\chi^2}{m_V^4}\,E_e^2\,(1+\cos\theta)\,,
\eeq{Msq_DP}
where $E_e$ and $\cos\theta$ are the electron energy and scattering angle, respectively, in the center-of-mass frame of the collision. Setting $\cos\theta=1$ (corresponding to $t=0$) in this equation yields Eq.~\leqn{c2DP}. }
 
During the kinetic decoupling and freeze-out process, the $\chi$ particles, 
to a good approximation, follow a Maxwell-Boltzmann distribution. Then,
 \beq
f_\chi =  \left( \frac{n_\chi}{n^{\rm eq}_\chi} \right)  f_{\chi}^{\rm eq}  
~~~~~\Longrightarrow~~~~~
 \rho_\chi = \left( \frac{n_\chi}{n^{\rm eq}_\chi} \right) \rho^{\rm eq}_\chi, ~~~~~  P_\chi = \left( \frac{n_\chi}{n^{\rm eq}_\chi} \right)P^{\rm eq}_\chi.
\eeq{MBrho}
Here `eq' denotes the values of the variables in chemical equilibrium, $\mu_\chi=0$:
\beqa
n^{\rm eq}_\chi &=& \frac{g_\chi m_\chi^2 T^\prime}{2\pi^2}K_2(m_\chi/T^\prime),\CR
\rho^{\rm eq}_\chi &=& \frac{g_\chi m_\chi^2 T^\prime}{2\pi^2}\left( m_\chi K_1(m/T^\prime) + 3T^\prime K_2(m_\chi/T^\prime)\right),\CR
P^{\rm eq}_\chi&=&\frac{g_\chi m_\chi^2 T^{\prime 2}}{2\pi^2}K_2(m_\chi/T^\prime),
\eeqa{eqvalues} 
and $n_\chi/n^{\rm eq}_\chi=e^{-\mu_\chi/T^\prime}$. The Boltzmann equations~\leqn{BEn},~\leqn{BErho} then reduce to a system of coupled partial differential equations for $T^\prime$ and $\mu_\chi$ (or equivalently $T^\prime$ and $n_\chi$). 

In the epoch of interest, the entropy of the universe is dominated by relativistic SM degrees of freedom: $s_0= \frac{2\pi^2}{45} g_{*s} T^3$, where $T$ is the SM plasma temperature, and $g_{*S}=10.75$ at the relevant temperatures ($0.5$ MeV $\lsim T \lsim 100$ MeV). The contribution of ELDERs to the entropy is suppressed both because they are non-relativistic, and because the number of degrees of freedom is small compared to SM. Neglecting this contribution, the time variable in the Boltzmann equations can be conveniently traded for the SM temperature: 
\beq
\frac{\partial}{\partial t} = -\left(1+ 3 \frac{T}{g_{*s}(T)} \frac{\partial g_{*s}(T)}{\partial T}\right)^{-1} HT \frac{\partial}{\partial T}.
\eeq{dt}     

\section{Kinetic Decoupling and Approximate Analytic Solution}

Since the relic density of ELDER dark matter is primarily determined at the time of its kinetic decoupling from the SM, we would like to obtain analytic insight into this process.  Kinetic decoupling occurs before freeze-out of the $3\to 2$ interactions, so that $\mu_\chi=0$ throughout the decoupling process. The ELDERs can then be completely characterized by their temperature $T^\prime$, whose evolution is dictated by Eq.~\leqn{BErho}. In this section, we describe an approximate analytic solution to this equation, which in turn yields an analytic estimate of the ELDER relic density. 

In the limit that the non-relativistic $\chi$ particles are scattering off thermalized relativistic $\psi$ particles, an approximate analytic form of the energy transfer rate integral, Eq.~\leqn{BEthermavg}, can be found. In Ref.~\cite{Bringmann:2006mu}, this was achieved by expanding the integrand in small momentum transfer. Here we present the necessary equations, but refer the reader to detailed calculation in the Appendix of \cite{Bringmann:2006mu}. First the thermally averaged energy transfer rate  is written in terms of the collision operator in the non-relativistic limit:
 \beqa
{n_{\chi}  n_{\rm \psi} }\langle  \sigma_{el} v \delta E\rangle &\simeq& \int d\Pi_{\chi_1} d\Pi_{\psi_1} d\Pi_{\chi_2} d\Pi_{\psi_2} (2\pi )^4\delta ^4\left(p_{\chi_1}+p_{\psi_1}-p_{\chi_2}-p_{\psi_2} \right)  \nn\\
&&~~~\times \left(\frac{p_{\chi_1}^2}{2 m_\chi}- \frac{p_{\chi_2}^2}{2 m_\chi}\right) f_{\chi_1} f_{\psi_1} \overline{\left|M_{\chi_1 \psi_1 \to \chi_2 \psi_2}\right|^2} \nn\\
&=&- \int d\Pi_{\chi_1} d\Pi_{\psi_1} d\Pi_{\chi_2} d\Pi_{\psi_2} (2\pi )^4\delta ^4\left(p_{\chi_1}+p_{\psi_1}-p_{\chi_2}-p_{\psi_2} \right)  \nn\\
&&~~~\times \frac{p_{\chi_1}^2}{2 m_\chi} ( f_{\chi_1} f_{\psi_1} - f_{\chi_2} f_{\psi_2}  ) \overline{\left|M_{\chi_1 \psi_1 \to \chi_2 \psi_2}\right|^2} \nn\\
&=& -\int d\Pi_{\chi_1} \frac{p_{\chi_1}^2}{m_\chi} C[f_{\chi_1}].
 \eeqa{thermcollision}
Using Eq.~(B.22) in \cite{Bringmann:2006mu}
 \beq
C[f_{\chi_1}] = \frac{g_\psi^2g_\chi}{12(2\pi)^3} m_\chi^2 c_n N_{n+3}^{\psi} \left(\frac{T}{m_\chi}\right)^{n+4} \left[ m_\chi T \nabla^2_{p_{\chi_1}} + \vec{p}_{\chi_1} \cdot \vec\nabla_{p_{\chi_1}} + 3 \right] f_{\chi_1 } ({p_{\chi_1}} )\,,
 \eeq{collisionterm}
where $c_n$ is the leading coefficient of the matrix element expanded in $E_\psi/m_\chi$ at zero momentum transfer 
\beq
  \mathop{\hspace{-6.5ex}\overline{\left|\mathcal{M}\right|^2}_{t=0}}_{\hspace{6.5ex}s=m_\chi^2+2m_\chi E_\psi}
 \equiv  c_n \left(\frac{E_{\psi}}{m_\chi}\right)^{n} + \ldots, 
\eeq{cn}
and
\beq
  N_j^\psi = \frac{j+1}{T^{j+1}} \int d E_\psi E^j_\psi f_\psi (E_\psi) = 
  \begin{cases} 
\left(1-2^{-j}\right)\,(j+1)!\,\zeta(j+1) & \psi \in \rm fermion, \\
(j+1)!\,\zeta(j+1) & \psi \in \rm boson .\\
   \end{cases}
\eeq{Ndef}
If expanding the matrix element around $t=0$ is not a good expansion, for instance, if the amplitude vanishes as $t\to0$, then one should replace \ref{cn} with the $t$-averaged matrix element~\cite{Binder:2016pnr}.
Taking  $f_\chi$ to be the Maxwell-Boltzmann distribution at temperature $T^\prime$ and integrating over the collision operator yields\beq
{n_{\chi}  n_{\rm \psi} }\langle  \sigma_{el} v \delta E\rangle \simeq  {n_{\chi}   }  \frac{c_n g_\psi^2 g_\chi m_\chi N^\psi_{3+n }}{32 \pi^3 }  \left(\frac{T}{m_\chi}\right)^{4+n}  (T^\prime-T).
\eeq{sigmavel}
 Note that when the two sectors have the same temperature, the energy transfer vanishes, which is expected for particles in thermal equilibrium. The energy transfer rate can be related to the more commonly used quantity $\langle \sigma_{ el} v \rangle$, as follows:
\beq
\langle\sigma_{el} v \delta E\rangle \simeq 2 (n+3) \frac{N^\psi_{3+n}}{N^\psi_{2+n}} \frac{T}{m_\chi} (T^\prime-T)\,\langle \sigma_{ el} v \rangle.
\eeq{Gscatter}

When the $3\to 2 $ process is active and the dark matter particles are non-relativistic, they follow equilibrium Maxwell-Boltzmann distributions, and the energy density Boltzmann equation~\leqn{BErho} gives a differential equation for the temperature 
\beq
\frac{\partial T^\prime}{\partial T} = 3 \frac{T^{\prime\,2}}{m_\chi T} +a\left(\frac{T}{m_\chi}\right)^{1+n} \frac{T^{\prime\,2}}{m_\chi^2}\frac{(T^\prime -T)}{m_\chi},
\eeq{dTpdT}
where 
\beq
a \equiv  \frac{c_n g_\psi^2 g_\chi m_\chi N^\psi_{3+n }}{32 \pi^3 H_{T=m_\chi}}.
\eeq{adef_app}
On the right-hand side, there are two competing terms. The first term contributes to the cannibalization of the dark matter, which tends to increase the dark temperature relative to the SM. The second term, which comes from the elastic scattering term, pushes $T^\prime \to T$. The scattering term falls faster with temperature, and at some point will no longer be able to compete. At that time, the dark matter will thermally decouple from the SM bath, and cannibalization will take over the evolution of the dark sector. This decoupling occurs roughly when the second term is of order one:\footnote{The kinetic decoupling temperature can also be estimated by observing that in equilibrium, the rate of energy transfer to the SM must keep up with the rate of kinetic energy release by $3\to 2$ annihilations: $n_e \langle\sigma_{el} v \delta E\rangle \sim -m_\chi^2 H T^{-1}$. According to Eq.~\leqn{Gscatter}, $\langle\sigma_{el} v \delta E\rangle \sim \langle \sigma_{ el} v \rangle T^2/m_\chi$. This approach, which was used in Ref.~\cite{Kuflik:2015isi}, gives a result consistent with Eq.~\leqn{Td}.}   
\beq
T_d \simeq m_\chi a^{-1/(4+n)}.
\eeq{td}
After decoupling, the second term can be ignored, and dark temperature grows only logarithmically relative to the SM temperature,
\beq
T^\prime \simeq \frac{T_d}{1 + 3 \frac{T_d}{m_\chi} \log \frac{T_d}{T} }\,,
\eeq{Tcann}
until the dark matter freezes out. 

We can attempt to find the analytic asymptotic behavior of the Boltzmann equations. Recasting Eq.~(\ref{dTpdT}) in terms of $x$ and $x^\prime$ yields
\beq
\frac{\partial x^\prime}{\partial x}=\frac{3}{x} + \frac{a x^{-n-4} (x-x^\prime)}{x^\prime}\,.
\eeq{xdifeq}
There appears to to be no closed form solution to the above differential equation, but the following differential equation does have a closed form solution:
\beq
\frac{\partial x^\prime}{\partial x}=\frac{3}{x} + \frac{a x^{-n-4} (x-x^\prime)}{x}\,.
\eeq{xdifeq2}
In the limit $x \ll x_d$, $x=x^\prime$ so the two differential equations are approximately the same. Likewise, when $x \gg x_d$, the 2nd term is negligible in both equations, so the change is not relevant. The closed form solution to Eq.~\leqn{xdifeq2} is
\beq
x^\prime = 
e^t \left(\left(\frac{a}{n+4}\right)^{\frac{1}{n+4}} \Gamma \left(\frac{n+3}{n+4},t\right)-\frac{3 \text{Ei}(-t)}{n+4}\right)
\eeq{xsol}
where $t \equiv \frac{a x^{-n-4}}{n+4}$, $\Gamma$ is the incomplete gamma function and $\text{Ei}(-t)=-\int_{t}^{\infty} \frac{e^{-z}}{z} dz$. The asymptotic limits of this solution at small and large $x$ are 
\beqa
x^\prime (x\to 0) &=& x \,,\\
x^\prime (x\to \infty) &=&  3 \log (x)+\left(\frac{a}{n+4}\right)^{\frac{1}{n+4}} \Gamma \left(\frac{n+3}{n+4}\right)-3 \log \left[e^{\frac{\gamma_E }{n+4}} \left(\frac{a}{n+4}\right)^{\frac{1}{n+4}}  \right].
\eeqa{xsolassym}
The second limit is very similar to the cannibalization result
\beq
\frac{\partial x^\prime}{\partial x}=\frac{3}{x} ,~~~~ x^\prime[x_d] = x_d  ~~~~~~  \Longrightarrow ~~~~~~ x^\prime = 3 \log (x)+x_d-3 \log (x_d ),
\eeq{xsolcann}
since $\Gamma (\frac{n+3}{n+4}) \approx e^{\frac{\gamma_E }{n+4}} \approx 1$. Therefore we make the identification for the decoupling temperature
\beq
x_d \simeq \left(\frac{a}{n+4}\right)^{\frac{1}{n+4}} \Gamma \left(\frac{n+3}{n+4}\right).
\eeq{xdanalytic}
This agrees with the rough estimate of Eq.~(\ref{td}), and provides the precise value of the numerical coefficient. Assuming instantaneous freeze-out of the $3\to 2$ annihilations at the dark sector temperature $x_f^\prime$ (corresponding to SM plasma temperature $x_f$), the dark matter yield is given by
\beq
Y(x_f) \equiv \frac{n^\prime (x_f)}{s (x_f)} = \frac{{g_\chi m_\chi^3 e^{x^\prime_f} }/{(2\pi x^\prime_f)^{3/2}} }{ \frac{2\pi^2}{45} g_{\star s} {m_\chi^3}/{x_f^3}} \simeq 
\frac{0.1 \frac{g_{\chi}}{g_{\star s,\, f}}a^{\frac{3}{8}} e^{-0.87 a^{1/4}} }{\left(1+\frac{3.4}{\sqrt[4]{a}}\log x_f\right)^{3/2}}.
\eeq{yf}
Here, the exponential dependence of relic density on the elastic scattering rate is manifest. We also note a logarithmic dependence on the temperature at freeze-out, which shows only a very minor dependence on the $3\to2$ rate, provided it is still active at decoupling. 

\section{Thermally-Averaged $3\to 2$ Rate}
\label{sec:rate}

Here we present the necessary formulas to calculate the thermally averaged $3\to2$ rate in thermal equilibrium, in the non-relativistic regime ($T\ll m_\chi$). Assuming Maxwell-Boltzmann distribution, which is justified when the dark matter is highly non-relativistic, the integral can be written in terms of 2-body and 3-body phase space integrals:    
 \beqa
\langle \sigma_{3\to 2} v^2 \rangle  &=& \frac{1}{3! 2!} \frac{1}{(n_\chi^{{\rm eq}})^3} \int d\Pi_{\chi_1} d\Pi_{\chi_2} d\Pi_{\chi_3} d\Pi_{\chi_4} d\Pi_{\chi_5} (2\pi )^4\delta ^4\left(p_{\chi_1}+p_{\chi_2}+p_{\chi_3}-p_{\chi_4}-p_{\chi_5} \right) \CR&&~~~\times f_{\chi_1} f_{\chi_2} f_{\chi_3}\overline{\left|\mathcal{M}\right|^2}\CR &=& 
\frac{1}{(n_\chi^{{\rm eq}})^3}\frac{g_\chi^5}{3! 2!} \int \frac{d^3 p_{\chi_1}}{(2\pi)^32 E_{\chi_1} }\frac{d^3 p_{\chi_2}}{(2\pi)^32 E_{\chi_2} } \frac{d^3 p_{\chi_3}}{(2\pi)^32 E_{\chi_3} }(2\pi)^4\delta ^4\left(p_{\chi_1}+p_{\chi_2}+p_{\chi_3}-p_0 \right) \nn\\
&&\times \frac{d^3 p_{\chi_4}}{(2\pi)^32 E_{\chi_4} }\frac{d^3 p_{\chi_5}}{(2\pi)^32 E_{\chi_5} }(2\pi )^4\delta ^4\left(p_0-p_{\chi_4}-p_{\chi_5} \right)  \frac{d^4 p_0}{(2\pi)^4}e^{E_0/T} \overline{\left|\mathcal{M}\right|^2}. 
\eeqa{C1}
Since we are interested in the case when the dark matter is non-relativistic, the system is approximately at rest. Therefore, to leading order, the  integrals can be performed in the center of mass frame.
The forms of the 3-body and 2-body space integrals are well known in this case:
\beq
\frac{1}{g_\chi^3}\int d\Pi_1d\Pi_2d\Pi_3 (2\pi)^4\delta^4(p_0 - p_1 - p_2 -p_3 ) = \frac{1}{(2 \pi)^3}\frac{1}{16 s} \int dm_{12}^2 dm_{23}^2\,,
\eeq{3body}
where $s = p_0^2$. The bounds of integration are
\beqa
m_{23,max}^2 &=& \frac{(s - m_\chi^2)^2}{4 m_{12}^2} - \frac{m_{12}^2}{4}\left(\lambda^{1/2}(m_{12}, m_\chi, m_\chi) - \lambda^{1/2}(m_{12}, m_\chi, \sqrt{s}) \right)^2 \,,\CR
m_{23,min}^2 &=& \frac{(s - m_\chi^2)^2}{4 m_{12}^2} - \frac{m_{12}^2}{4}\left(\lambda^{1/2}(m_{12}, m_\chi, m_\chi) + \lambda^{1/2}(m_{12}, m_\chi, \sqrt{s}) \right)^2 \,, \CR
m_{12,max}^2 &=& (\sqrt{s} - m_\chi)^2 \,,\CR
 m_{12,min}^2 &=& 4m_\chi^2\,,
\eeqa{bounds}
where  $\lambda(x,y,z) = \left(1- (z+y)^2/x^2 \right)\left(1- (z-y)^2/x^2 \right)$. The 2-body phase-space integral is
\beq
\frac{1}{g_\chi^2}\int d\Pi_4d\Pi_5 (2\pi)^4\delta^4(p_0 - p_4 - p_5 ) = \frac{1}{8\pi} \lambda^{1/2}(\sqrt{s}, m_\chi, m_\chi).
\eeq{2body}
Finally, the remaining $p_0$ integral can be written as 
\beq
\dfrac{d^4p}{(2\pi)^4} = \frac{1}{(2\pi)^3} \int^\infty_{9m_\chi^2} ds \int^\infty_{\sqrt{s}} dE_0 \sqrt{E_0^2-s}\,.
\eeq{p0}
Putting everything together, we obtain
\beq
\langle \sigma_{3\to 2} v^2 \rangle = \frac{g_\chi^5}{(n_\chi^{{\rm eq}})^3}\frac{1}{768} \frac{1}{(2\pi)^7} \int_{9m_\chi^2}^\infty \frac{ds}{s} \sqrt{1- \frac{4 m_\chi^2}{s}} \, \int_{\sqrt{s}}^\infty 
dE_0 e^{E_0/T} \sqrt{E_0^2-s} \, \int dm_{12}^2 dm_{23}^2 \overline{\left|\mathcal{M}\right|^2}.
\eeq{23inter}
If the matrix element has significant dependence on kinematics even in the non-relativistic regime, the remaining integrals have to be evaluated numerically. This is the case in the Choi-Lee model of Section~\ref{sec:CL}, where a resonance at $\sqrt{s}\approx 3m_\chi$ can lead to rapid change of the matrix element with $s$ near threshold. Our analysis of that model is therefore based on numerical evaluation of Eq.~\leqn{23inter}. In most cases, however, the matrix element in the non-relativistic regime can be approximated as a constant, independent of kinematics. In this case, all integrals in Eq.~\leqn{23inter} can be evaluated analytically. This yields
\beq
\langle \sigma_{3\to 2} v^2 \rangle = \frac{\sqrt{5} g_\chi^2}{2304 \pi m_\chi^3}  \overline{\left|\mathcal{M}\right|^2}.
\eeq{23final}

\bibliographystyle{jhep}
\bibliography{lit}\bibliographystyle{jhep}

\end{document}